\documentclass[sigconf,nonacm]{acmart}

\usepackage{soul}
\usepackage{subfigure}
\usepackage{algorithm}
\usepackage{algorithmic}
\usepackage{multirow}
\usepackage{multicol}
\usepackage{enumerate}
\usepackage{amsmath}

\usepackage[most]{tcolorbox}
\definecolor{MidnightBlue}{HTML}{006895}

\usepackage{pifont}
\newcommand{\cmark}{\ding{51}}%
\newcommand{\xmark}{\ding{55}}%

\usepackage{xspace}
\usepackage{graphicx}

\AtBeginDocument{%
  }

\newcommand*{\eg}{e.g.\@\xspace}
\newcommand*{\ie}{i.e.\@\xspace}
\newcommand*{\etal}{et al.\@\xspace}

\newcommand{\SystemName}{\textsc{Explorer}\xspace}

\begin{document}

\title[\SystemName]{\SystemName: Robust Collection of Interactable GUI Elements}

\author{Iason Chaimalas}
\authornote{Both authors contributed equally to this research.}
\email{iason.chaimalas.20@ucl.ac.uk}
\orcid{0009-0008-1460-7678}
\affiliation{%
  \institution{University College London}
  \city{London}
  \country{United Kingdom}
}

\author{Arnas Vy\v{s}niauskas}
\authornotemark[1]
\email{arnas.vyniauskas.20@ucl.ac.uk}
\affiliation{%
  \institution{University College London}
  \city{London}
  \country{United Kingdom}
}

\author{Gabriel Brostow}
\email{brostow@cs.ucl.ac.uk}
\affiliation{%
  \institution{University College London}
  \city{London}
  \country{United Kingdom}}

\renewcommand{\shorttitle}{\SystemName: Robust Collection of Interactable GUI Elements}
\renewcommand{\shortauthors}{Chaimalas~\etal}

\begin{abstract}

    Automation of existing Graphical User Interfaces (GUIs) is important but hard to achieve. Upstream of making the GUI user-accessible or somehow scriptable, even the data-collection to understand the original interface poses significant challenges. For example, large quantities of general UI data seem helpful for training general machine learning (ML) models, but accessibility for each person can hinge on the ML's precision on a specific app. 

    We therefore take the perspective that a given user needs confidence, that the relevant UI elements are being detected correctly throughout one app or digital environment. We mostly assume that the target application is known in advance, so that data collection and ML-training can be personalized for the test-time target domain. The proposed \SystemName system focuses on detecting on-screen buttons and text-entry fields, \ie interactables, where the training process has access to a live version of the application. The live application can run on almost any popular platform except iOS phones, and the collection is especially streamlined for Android phones or for desktop Chrome browsers.
    \SystemName also enables the recording of interactive user sessions, and subsequent mapping of how such sessions overlap and sometimes loop back to similar states. We show how having such a map enables a kind of path planning through the GUI, letting a user issue audio commands to get to their destination. Critically, we are releasing our code for \SystemName openly at {\color{magenta}{\href{https://github.com/varnelis/Explorer}{\texttt{https://github.com/varnelis/Explorer}}}}.
  
\end{abstract}


\maketitle

\section{Introduction}
We focus on the process of collecting \emph{relevant} data for AI automation of graphic user interfaces (GUIs). There is now significant appetite to build computer ``bots'' that can use, or at least understand, the GUIs of existing software. Getting a computer to use computers is far from just a philosophical exercise. Improved screen-readers \cite{ScreenRecognitionScreenReaderAppleCHI21} are one successful commercialized example, where blind and low-vision users benefit from an AI-driven bot as the intermediary to their laptop or phone. Additionally, there is increasing public awareness and demand for personalized GUI automators, as exemplified by the record sales of the Rabbit R1 gadget, despite its flaws~\cite{verge_rabbit_r1_2024}. Programming-by-Demonstration~\cite{watchwhatido,wishcommand} (PbD) systems seek to sufficiently understand a GUI's widgets to be able to repeat chains of events, such as needed for automatic software testing on device farms~\cite{Su2021SoftwareGUItesting}. We are especially motivated by downstream use-cases where someone is building or benchmarking a personalized AI assistant, or adapting existing GUIs for accessibility. 

There is a small but growing community of researchers and developers in this space~\cite{JiangExtendedAbstrCHI23, JiangCHI22extendedAbstract}. They train Machine Learning (ML) models on supervised data to make software capable of some form of GUI understanding. Rico~\cite{rico} and Wu~\etal's Never-ending UI Learner~\cite{neverending} are exemplary proofs of concept, because they show how much can be achieved with, respectively, hand-validated or auto-crawled datasets. But they also demonstrate the acute need for the proposed \SystemName, our open codebase of data-collection, self-labeling and GUI automation tools for both desktop and mobile (Android) GUIs. RICO~\cite{rico} is a fixed-size snapshot of GUIs from 2017. It would cost \$20k to recreate and more to grow it now. Indeed, NeverEnding~\cite{neverending} demonstrates how effective auto-collection (``crawling'') can be in the mobile iOS ecosystem. They collect labeled pairs from $6,461$ distinct apps for $5000$ hours. But neither the code nor the data from NeverEnding is available now. The same is true for data described by Spotlight~\cite{spotlight}. There could easily be commercial impediments keeping Apple and Google from sharing the code and data in these publications. We do not have such limitations, so our contributions are:
    \begin{itemize}
        \item A suite of data-collection tools for automatically finding and labeling ``clickable'' elements, effective, with some differences, on Windows, Mac, and Linux desktops, and Android phones, 
        \item A model for recording, mapping, and then using application-specific user sessions.
    \end{itemize}

Both contributions are proposed as complementary to existing works, and our code and non-personal data are being released on GitHub and Hugging Face, on acceptance. Our code for scalable data collection builds on the Kotlin~\cite{KotlinDocs} and Selenium~\cite{selenium} APIs. The second contribution focuses on letting users enroll and then automatically access one application deeply and reliably, with obvious trade-offs compared to automatically detecting interactables fairly well across numerous apps. We model the different states when using a GUI-based application as an interconnected graph of unique states, where user actions (\eg text input, mouse clicks, phone taps) induce state transitions. Our \SystemName's architecture involves ML object detection and screen similarity, to understand and actuate GUIs with no reliance on platform-specific APIs, although those were helpful for prior PbD works~\cite{sugilite,pumice,etna,vasta,actionshot}. As a specific proof of concept, our \SystemName enables a user to navigate a GUI with hands-free voice commands, and to replicate an end-to-end user-demonstrated GUI traversal.

We evaluated the \SystemName in a variety of ways, and on multiple popular computer websites, namely KhanAcademy, Wolfram Alpha, Wikipedia, and phone applications, \ie Spotify and the Android Telephone app. Perhaps surprisingly, our experiments indicate that higher accuracy hinges on more than general-purpose large-scale data collection.

\section{Related Work}

    In the field of GUI Automation, emphasis must be placed on ease of use for the end-user, and applicability to a wide task set. Generally, while application-specific internal tools like SDKs and APIs enable automation on a per-application basis, they require considerable programming by experienced users. The same limitation applies to early systems \textit{Sikuli}~\cite{sikuli} and \textit{Prefab}~\cite{prefab}, which detected and reverse-engineered GUI widgets (\eg buttons, toolbars, etc.) based on exact pixel-appearance. Concurrently, \textit{Koala}~\cite{koala} and \textit{CoScripter}~\cite{coscripter} automated Firefox-based tasks executed by a user, using a browser-specific plug-in. These methods enabled a more abstracted control of the GUI environment, but still required end-user programming. Consequently, low-code \cite{lowcode} and no-code~\cite{sikuli-slides} automation solutions have arisen within the PbD paradigm, and the concept of a ``demonstration'' has advanced to a continuous video recording of the GUI screen as the user executes a specific action sequence\footnote{This involves keyboard or mouse events in computers, and finger-taps in phones \& tablets}~\cite{sikuli-slides}.
    
    To detect user-actuated interactables on Android devices, \textit{VASTA} \cite{vasta} and \textit{V2S}~\cite{v2s} used Computer Vision object detectors~\cite{retinanet,yolo,fasterrcnn,fcos}, which is more platform-independent than early systems using pixel-detection \cite{prefab,sikuli}, user input \cite{hilc-long,hilc-short}, APIs or other internal tools \cite{actionshot,etna,sugilite,pumice, webgum}. Data for most of these models has been sourced via expensive and laborious crowdsourcing~\cite{vins, screenrel}, which is also prone to human error~\cite{tapShoe,humanerror}. For instance, in a Screen Similarity study for Android phones by Feiz~\etal~\cite{screenrel}, a Quality Assurance (QA) team reviewed the study's dataset manually labeled by 36 human crowd workers; the QAs disagreed with the original crowd workers in 65.68\% of cases that different-labeled groups were truly different GUI states. Meanwhile, recent works that automate the data sourcing are often noisy~\cite{webui} or limited to simulation-based Atari environments~\cite{world-of-bits, miniwob++}. An alternative to interactable detector models was proposed by \textit{SeeClick} \cite{cheng2024seeclick} -- by using Large Vision-Language Models (LVLMs), they were able to find text, widgets and icons using user prompts; for this they utilize a large 9.6B parameter model, and achieved 78\% accuracy on mobile-text benchmark, but only 30\% on desktop-icon/widget benchmark.
    
    In the field of PbD, for identifying whether two screens are similar, existing works use \textit{Screen Similarity} models; they involve a Siamese architecture~\cite{dimredux_siamese,similaritymetric} that learns a latent embedding space where same-label inputs are closer than different-label inputs in terms of a given distance metric. In \textit{WebUI}~\cite{webui}, the aim was to determine if two screenshots come from the same application or webpage. Similarly, Feiz~\etal~\cite{screenrel} seek to determine when two screens on a phone differ meaningfully. Moreover, they have a secondary module to detect scrolling within the same GUI state. \textit{NeverEnding}~\cite{neverending} uses screen similarity for deciding the crawling strategy in navigating a phone device and live fine-tuning their interactable detector. In contrast, we propose to use this model in end-to-end applications for robust understanding of GUI state changes. This allows us to dynamically identify screen loading after a state transition and robustly identify interactables selected by a user during a task demonstration; meanwhile, existing systems largely rely on fixed timers~\cite{hilc-long}, APIs~\cite{etna}, and exact logged user data~\cite{v2s}.
    
    We follow the trend of publicly sharing data \cite{webui, cheng2024seeclick, webgum} and we provide not only datasets, but also source code for (1) data-collection on most modern Operating Systems (Windows, MacOS, Android), and (2) end-to-end GUI automation\footnote{\SystemName repository at {\color{magenta}{\href{https://github.com/varnelis/Explorer}{\texttt{https://github.com/varnelis/Explorer}}}}; Android-specific data collection at {\color{magenta}{\href{https://github.com/IasonC/AndroidGUICollection}{\texttt{https://github.com/IasonC/AndroidGUICollection}}}}.}

\section{\SystemName for Environment Exploration and Data Collection}\label{sec:Scraping}
\begin{figure}[h!]
        \centering
        \includegraphics[width=\linewidth]{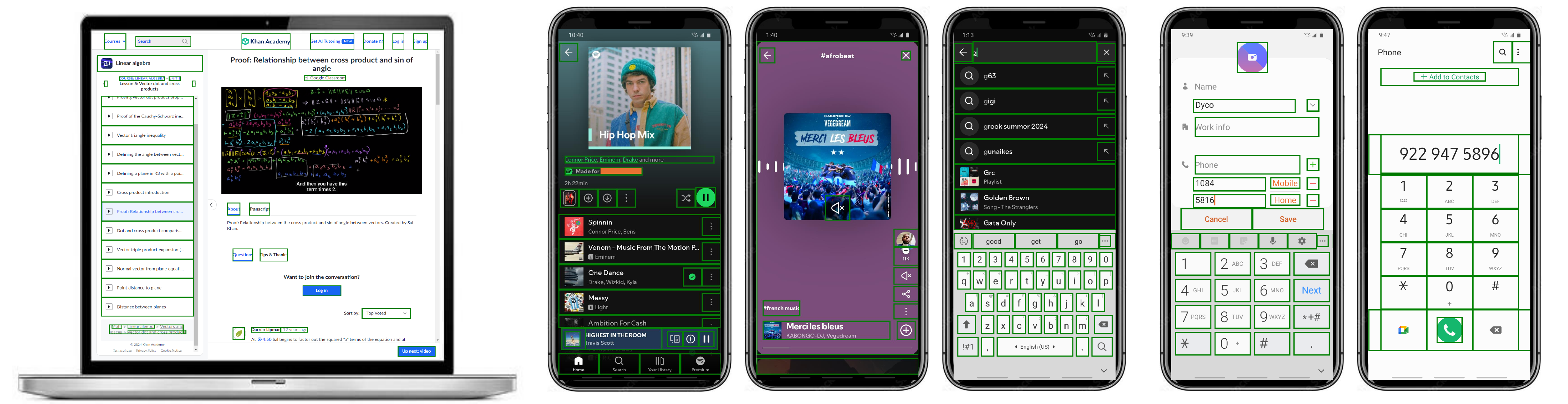}
        \caption{Visualization of data collection and auto-labeling, with ground-truth bounding boxes in \textit{green}. Personal data is covered (orange).}
        \label{fig:data-interact}
    \end{figure}

    The \SystemName system comprizes multiple components. The main groupings are data collection, described in the rest of this section, and subsystems for GUI automation, covered in Sec.~\ref{section:Methods_GUI}. For our purposes, success in data collection means being able to intentionally capture imagery and basic interaction events covering an app's different modes and screens. Specifically, we want to capture pressing of interactables, \eg buttons and text entry boxes, and to do so for a broad range of GUI platforms. Examples are shown in Figure~\ref{fig:data-interact}. We also capture information useful for assessing screen similarity. 

    We propose three modes of data collection. The first is a generalized platform-independent algorithm for collection and labeling that leverages visual screen changes from hover effects on GUI elements. The other two are designed for speed instead of generality. One is for Chrome browsers in Windows, MacOS, and Linux, and will be demonstrated on KhanAcademy, Wolfram Alpha, and Wikipedia. The other focuses on phone apps in Android, demonstrated on Spotify and the built-in Telephone app.

    \begin{figure}[h!]
        \centering
        \includegraphics[width=\linewidth]{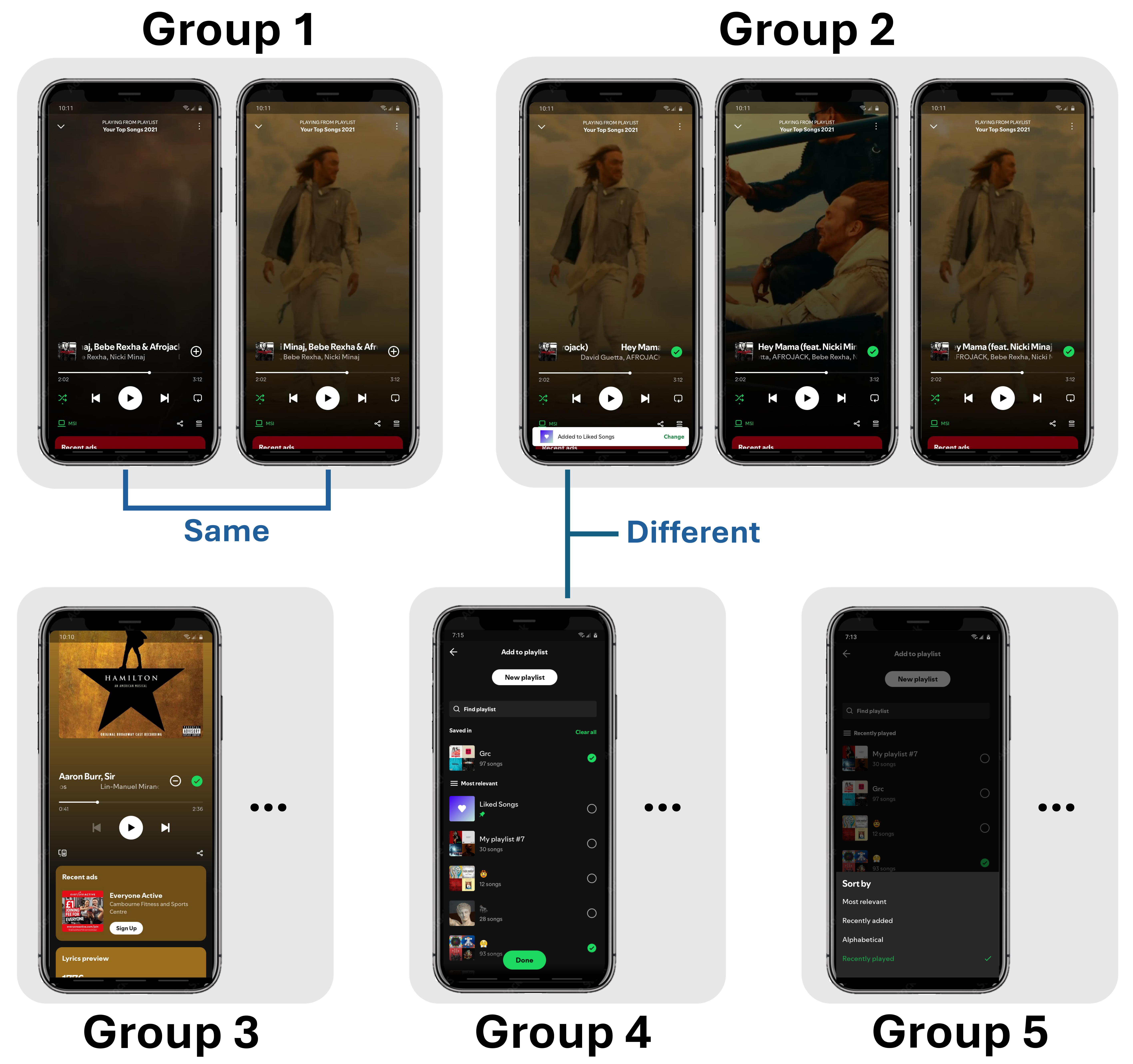}
        \caption{Visualization of data labeling for the Screen Similarity task, with same-state GUI screens labeled in the same group. Screenshots labeled to different groups (e.g. Group 2 and 4) are different GUI states. Hence, note that training labels (``same'' or ``different'' state) for the Siamese network can be inferred based on group membership.}
        \label{fig:data-screensim}
    \end{figure}

    \subsection{Generalized GUI Data Collection}
    \label{section:datacollect}
        
\begin{figure}[h!]
            \centering
            \includegraphics[width=0.95\columnwidth]{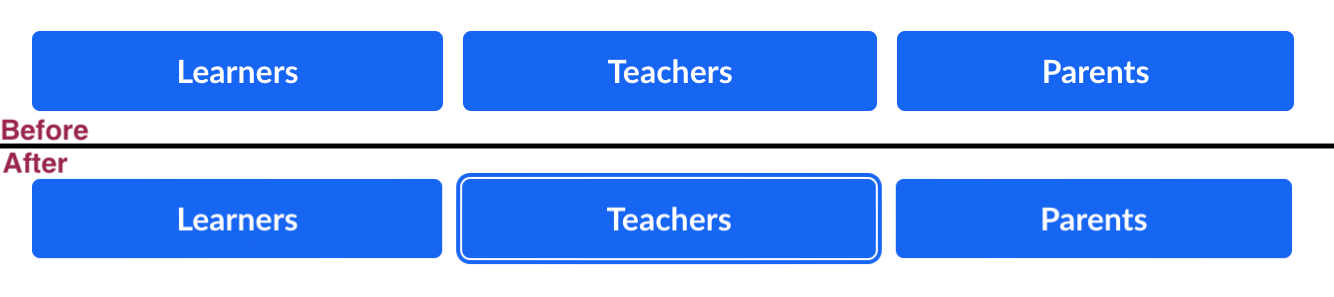}
            \caption{Visual change induced \textit{after} hovering over the ``\textit{Teachers}'' interactable with a mouse, on the \textit{KhanAcademy} home website. Similar changes occur for desktop-computer GUI elements, either on or sometimes around the interactable, \eg tooltips.}
            \label{fig:hovering-before-after}
         \end{figure}

        We assume that hovering over the interactables with a mouse activates them by changing their appearance, or providing other visual indications to the user that they are clickable. This assumption is true for most modern GUIs. Capturing sequences of screenshots reveals what changes occur before and after the mouse movement (Figure~\ref{fig:hovering-before-after}), and enables us to compute the pixel-level \textit{difference image}. Non-zero differences indicate that a specific mouse position as an interactable. Each interactable usually renders a specific and reproducible change. Hypothetically, by observing each possible mouse position, every pixel could be assigned to a specific interactable. This is particularly slow, because it requires some time gaps for the GUI to update, before checking the next mouse position.
                
        The number of pixel checks is adjustable to meet speed or cost requirements.
        Interactables generally have simple shapes, and the parameter $h$ defines the minimal possible width of interactables when collecting a new dataset. We can assume that two pixels at distance $\leq h$ from each other are neighbors, thus reducing the run time for a coarse search by a factor of $h^2$. Our algorithm is analogous to the optimized strategy for locating enemy ships in the \textit{BattleShip} table-top game~\cite{rodin1988developing} for 2-dimensional ships. After gathering the coarse view of the screen, we then interpolate by applying rule-based checking inspired by cellular automata~\cite{cellular-automata}. This method decides whether to check each interpolated pixel or assume its type by looking at the types of the neighbouring pixels.
    
        Finally, we group pixels together into interactables by inferring the group from the hash of the difference image: pixels belonging to the same interactable will have the same difference image, and hence, the same hash. We typically bound the group of pixels with a bounding box, but this is optional.

        \paragraph{Limitations.}
        This generalized algorithm depends only on control over the on-screen mouse position and access to GUI screenshots. Using remote access software such as TeamViewer, this can run at scale for almost all platforms, devices, or OSs. Wu~\etal~\cite{neverending} used VNC. However, a basic assumption is that the GUI environment visually reacts to a hover event over an interactable. That is not natively supported by some phones. Further, with dense sampling or small $h$, this generalized algorithm brings long run-times. %
        The next sections offer specialized tools for more efficient data collection in two specific scenarios: websites and Android devices.
    
    \subsection{Website Data Collection}
    \label{section:webdata}

        We extend the \textit{Selenium}~\cite{selenium} framework to programmatically perform the same kind of data-collection, but more efficiently, on webpages. This works for websites when viewing through computer-based web-browsers that use the Chrome libraries (\eg Chrome, Microsoft Edge, Firefox, Opera, etc.). To label ground-truth bounding boxes for Interactable Detection, we parse the extracted HTML \texttt{XPath}\footnote{\texttt{XPath} attributes are coded descriptors for on-screen elements} elements (\eg buttons, inputs), to identify the on-screen coordinates of interactable elements. We filter invalid interactables via the following stages:
        \begin{enumerate}
            \item small interactables below a minimum pixel area,
            \item interactables that are set to be invisible by the page itself, and
            \item interactables that are positioned outside their \texttt{XPath} parent element\footnote{Interactable elements are described by their \texttt{XPath} in HTML, and they are structured as a dependency tree; smaller interactables depend on a \textit{parent} interactable -- \eg the ``top toolbar'' is often the parent of a \texttt{Menu} and \texttt{Homescreen} button in many websites.}.
        \end{enumerate}

        To access a \emph{relevant} list of URLs to crawl and label, we employ two methods:
        \begin{enumerate}
            \item \ul{Web Tree Hierarchy}: We crawl a sub-graph of webpages representative of the total page distribution of an entire website, by inspecting the top-level website HTML. We then sample without replacement sub-graph webpages (GUI states) from each top-level URL group (\eg \texttt{/search}, \texttt{/about}).
            \item \ul{Simulated Human Traversal}: After recording one GUI state, we induce a random user action (click, scroll) to cause a state transition to a new GUI state.
        \end{enumerate}
        Website sub-graph crawling techniques (Method 1), also deployed by Wu~\etal~\cite{webui}, are best used for websites where pages can be strictly categorized based on appearance, and category size does not represent the popularity of that category (\eg KhanAcademy). However, Method 2 is more natural and better reflects the distribution of GUI states encountered by a real user. It fits better for websites where certain visual attributes are common throughout the website sub-graph (\eg Wikipedia). We apply Method 1 to the \textit{KhanAcademy} website, and Method 2 to \textit{Wolfram Alpha} and \textit{Wikipedia}. Both methods were deployed in Windows and MacOS computers, but are also trivially applicable to \textit{Linux}. %

        Finally, in terms of training data for Screen Similarity, Wu~\etal have also gathered an automated dataset for computer websites~\cite{webui}. However, we adopt a stricter interpretation of GUI state: while WebUI differentiates GUIs belonging to different websites, we further distinguish GUI states separated by user actions (left click, scroll, keyboard/text injection) even within the same \textit{webpage}. For Screen Similarity data, we modify our Methods 1 and 2, such that at each GUI state we simulate various events that do not change the state (\eg playing on-screen videos at various timestamps, hovering over interactables). For all such events, we capture screenshots and label them within the same data group (Figure~\ref{fig:data-screensim}).
        
    \subsection{Android Data Collection}
        To collect data from Android devices more efficiently, we developed a custom application programmed in \texttt{Kotlin}, available in {\color{magenta}{\href{https://github.com/IasonC/AndroidGUICollection}{\texttt{GitHub}}}}.
        Our app requests read-write permissions for the phone, and launches a custom Android Accessibility Service. This Service listens for user actions (tap/click, scroll) on the phone screen, and waits for a fixed timer (empirically set at 4 seconds). It then launches an Interruptible Service Routine (ISR) in a parallel thread to capture a high-resolution screenshot, and updates data in memory as follows:
        \begin{enumerate}
            \item \ul{Interactable Detection}: Writes a key-value pair for the current GUI state associated with a randomly-generated UUID, as \texttt{\{uuid: List[bbox]\}}; the list of bounding boxes for valid interactables is obtained by extracting and traversing the Accessibility Tree of the currently-visible app.
            \item \ul{Screen Similarity}: Gathers multiple screenshots at 2-second intervals; if a video/song is currently playing (\eg \textit{Spotify} app), the timestamp is altered within each interval. All screenshots are defined as the same GUI state, and are written in memory as \texttt{\{``group-uuid'': List[uuid]\}}.
        \end{enumerate}
        Finally, after data has been gathered at a given GUI state, our algorithm causes a state transition via a random choice among common user actions: (i) click on random interactable, (ii) scroll down, or (iii) text injection in text-boxes with a random text string. %
    
        \paragraph{Design Choices and Compatibility} We have developed our Android data collection app to comply with programming design and security requirements set for Android API 29 and higher. For instance, the screenshot-grabbing ISR is launched as a \texttt{Foreground Service} with one-time permissions requested from the device user. Further, Android Accessibility Services in API 29+ cannot programmatically click on keyboard elements due to security constraints. Instead, we simulate keyboard actions by filtering elements of the Accessibility Tree for text-box attributes and directly injecting text. Meanwhile, our app is also backward-compatible with lower API levels, covering the majority of modern Android devices.

\begin{figure}
            \centering
            \includegraphics[width=0.9\linewidth]{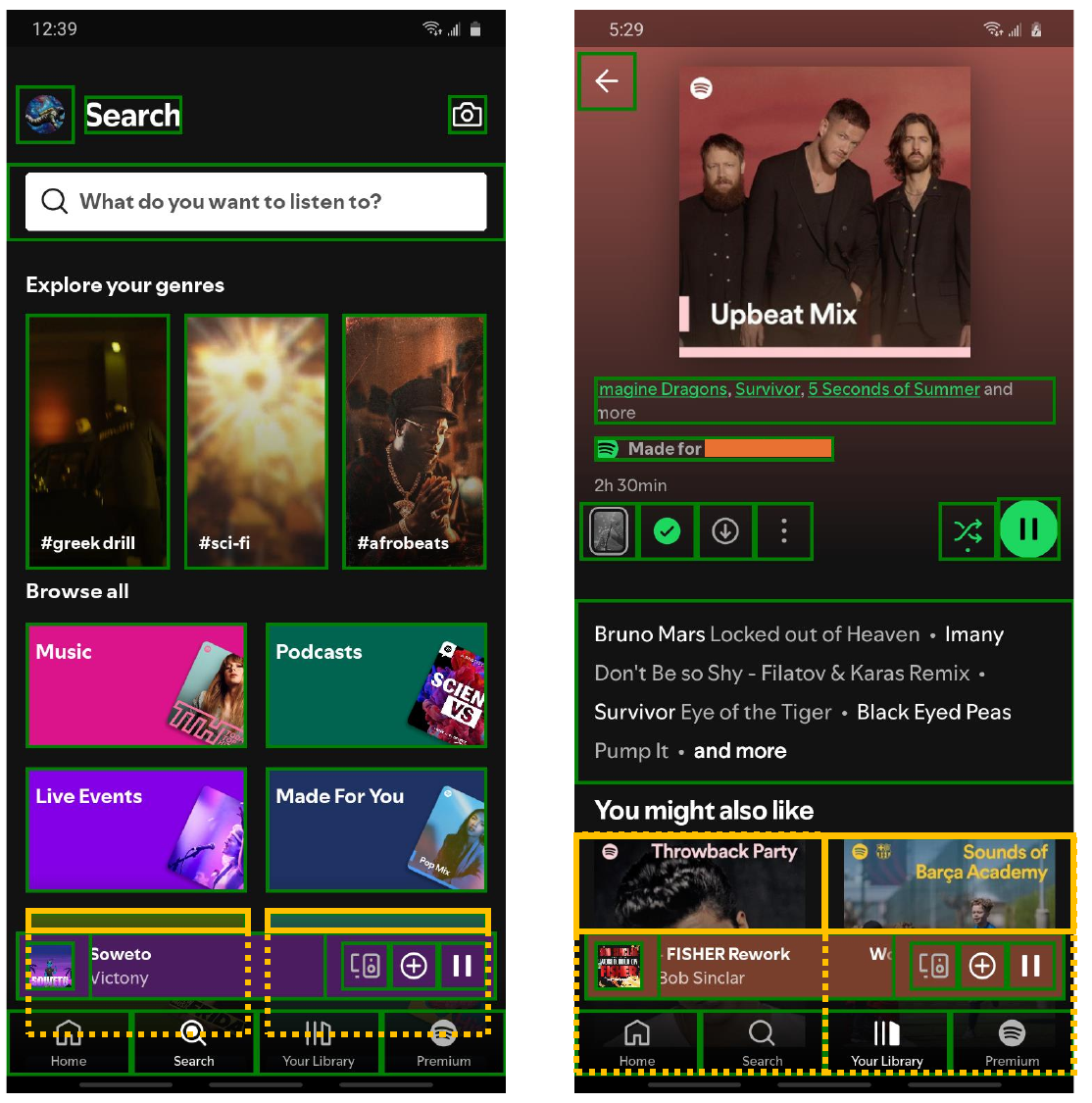}
            \caption{Examples of two common occlusions in the Android \textit{Spotify} app. Occluded interactables' \texttt{bbox} retrieved from the Android Accessibility Tree is illustrated with yellow dashed-lines. Compare this to the truncated solid-yellow \texttt{bbox} above the bottom song-banner, which is the true size shape of that interactable. Personal information is covered (orange).}
            \label{fig:android-truncation}
        \end{figure}

        Another important design choice relates to handling occlusions among interactables. While \textit{Selenium} captures z-order, none of the web-sites we tested use it. Also, the Accessibility Tree extracted from the currently-visible Android app does not explicitly indicate vertical hierarchy (``Z-order''), nor does it truncate the bounding box of interactables occluded by higher-Z elements. Therefore, cases arise where multiple interactable bounding boxes are given in a pixel position where only the top-layer interactable is truly clickable. See Figure~\ref{fig:android-truncation}. We assume that the Accessibility Tree is heuristically structured such that child elements (left-to-right) are higher-Z than parents, which is true in most modern apps. Next, we traverse the Accessibility Tree with Breadth-First Search to maintain Z-order, and identify interactable occlusions. Heuristically, if the overlapping interactables do not share a descendant-ancestor relation, the lower-Z interactable is truncated.

        \paragraph{Novelty and Extensions} Our Android automated data-gathering app stands out from previous Android-focused efforts. 
        Unlike prior crowd-sourced datasets~\cite{vins,rico,enrico,screenrel}, our data-gathering process is completely automated beyond the initial one-time permissions granted by a physical user. 
        Moreover, unlike the automated data-gathering by Wu~\etal, who simulate a phone resolution of computer-based webpages~\cite{webui}, our app can gather data in both websites and non-website generalized apps directly on a physical Android device. 
        Further, like ~\cite{neverending} is for iOS, our Android data collection is scalable and implementable in edge-hardware like Android CPUs.
        Thus, we can achieve similar data-sizes to large open-source datasets~\cite{vins,rico,enrico} within a few days, with very high labeling accuracy. Future work may extend automated data collection to iOS devices, so such data could be publicly available also. This would involve a similar programming model to our current Kotlin app for Android, but written instead in Swift or Objective-C, and targeting iOS Accessibility APIs.

    \subsection{Data for our Experiments}
    \label{section:dataacq}

        We have deployed our website and Android data collection algorithms to automatically gather high-quality training data for our \SystemName system's Interactable Detection and Screen Similarity subsystems. Specifically, Table~\ref{tab:datasources} lists the GUI applications that we collected, which cover a range of educational, professional, and entertainment use-cases.
        \begin{table*}[h!]
          \caption{GUI Applications collected for \SystemName training data. Note the shorthand terms ``Interact'' and ``ScreenSim'' for Interactable Detection and Screen Similarity tasks, respectively.}
          \label{tab:datasources}
          \begin{tabular}{cccccc}
            \toprule
            \multirow{2}{*}{GUI Application} & \multirow{2}{*}{Type} & \multirow{2}{*}{Platform} & \multirow{2}{*}{Resolution} & \multicolumn{2}{c}{Data Size} \\
            & & & & Interact & ScreenSim \\
            \midrule
            \textit{KhanAcademy} & Website & Windows & $1249 \times 1234$ & $2841$ & $378$ \\
            \textit{Wolfram Alpha} & Website & MacOS & $2400\times 1504$ & $1109$ & $-$ \\
            \textit{Wikipedia} & Website & MacOS & $2400\times 1398$ & $1086$ & $-$ \\
            \textit{Spotify} & App & Android & $720 \times 1560$ & $1207$ & $451$ \\
            \textit{Telephone \& Contacts} & App & Android & $720 \times 1560$ & $525$ & $-$ \\
          \bottomrule
        \end{tabular}
        \end{table*}
        We note that these gathered datasets are an order of magnitude smaller than established PbD datasets~\cite{vins,rico,enrico,webui,screenrel}, despite our capability to scalably collect equivalent quantity in reasonable time. Our proof-of-concept shows that through high-quality collection of target-domain data and our ML architecture, we can match or even beat more generalized models. 
        Further, we collect Interactable datasets across multiple GUI platforms, types and screen resolutions, to demonstrate the scalability, robustness, and applicability of our data-gathering method. 

        We commit to release the web-based datasets on \textit{KhanAcademy}, \textit{Wolfram Alpha} and \textit{Wikipedia}. We cannot release data on the two Android applications, as those contain personal information. However, our open data collection code for Android allows everyone to easily gather extensive data.

\section{\SystemName System for GUI Automation}
\label{section:Methods_GUI}

    We address the prevalent generalization and robustness issues in current end-to-end GUI Automators with the \SystemName system. The \SystemName is a Supervized agent trained to \textit{visually understand} GUIs and actuate correct interactable elements in inference-time, using solely the visual GUI screen and low-level system actions (\eg key-presses, mouse clicks, phone-screen taps) as inputs. We achieve this using a pure Computer Vision pipeline based on our \textit{Interactable Detector}, \textit{Screen Similarity} and\textit{ Action Matching }ML Subsystems.
    
    For the \SystemName to have a more consistent understanding of the computer system, we use a novel approach of structuring how the system understands the screen. Instead of the \SystemName constantly looking for target interactables, our program waits for a \textit{steady state}. We demarcate a general GUI environment into distinct states which change based on some user actions or various computer actions (\eg update window pop-up). In most cases, after a user performs an action, the system cycles through many loading states and settles on the next expected state. Just like a human, the \SystemName waits for a consecutive stream of frames that are similar. If adjacent frames are considered different by the \textit{Screen Similarity} model, the state is assumed to be changing until eventually the system reaches steady-state again. This eliminates uncertainty introduced by network- or system-loading times. Moreover, it better supports the recording format of a user demonstration as a \textit{trace} of discrete GUI states, by keeping it synchronized with the user's system. Therefore, we are able to convert a very dynamic computer system into a set of discrete states separated by expected uncertainty-frames, which we can ignore. 

    Utilizing this strategy, we demonstrate the \SystemName's capabilities in (1) \textbf{traversing the GUI environment} according to hands-free end-user spoken commands, and (2) \textbf{automating GUI actions} in a \textbf{platform-independent} GUI environment with no platform-specific API dependence and robustness to GUI changes during automated execution. Further, our novel architecture presents a promising direction for multi-modal large-scale GUI agents and online learning.
    
    \subsection{Subsystem 1: Interactable Detector Model}
    \label{section:interactabledetector}
        Our \textit{Interactable Detector} model is an FCOS network~\cite{fcos} trained separately for each target GUI application with our respective labeled dataset in Table~\ref{tab:datasources}. The FCOS architecture involves a back-bone ResNet model with feature maps C3-C5 in decreasing dimensionality; the back-bone C3-C5 is fed to a Feature Pyramid Network (FPN)~\cite{fpn} with feature maps P3-P7, which identify salient features spanning fine-grained (P3) to large (P7) areas. Further, \textit{Centerness} feature maps Ctr3-Ctr7 are computed, emphasizing the location of target objects (interactables here) on each corresponding P3-P7. Maps P3-P7 and Ctr3-Ctr7 contribute to the final bounding-box prediction for target interactables on an input GUI screenshot.

        We choose FCOS because it is an \textit{anchor-free} object detector, unlike RetinaNet~\cite{retinanet}, Faster R-CNN~\cite{fasterrcnn} and early versions of YOLO~\cite{yolo}. Since the size of interactables may vary across multiple applications, websites, or general OS platforms, initial bounding-box predictions from fixed-size anchor boxes may be brittle and application-specific. Instead, FCOS yields a generalisable model for GUI screenshots with satisfactory inference speed (FPS).
        
        Among existing works, Wu~\etal also deployed an FCOS for interactable detection~\cite{webui}. They pre-trained the model on their automatically-crawled dataset of website screenshots at various dataset sizes, including 350k, 70k, and 7k webpages. Wu~\etal then fine-tuned on the public VINS dataset~\cite{vins} of $4800$ phone screens. We focused on the pre-processed and cleaned set of 7k websites (\textit{WebUI-7k} dataset), as it is the fastest to train while leading to a downstream VINS-finetuned-performance almost on-par with the 350k dataset. For our Interactable Detector, we explore both training from scratch on our target-GUI datasets or fine-tuning fromWebUI 7k or VINS data ($\S$\ref{section:interactable-eval}).

        To match the training configuration of Wu~\etal~\cite{webui}, we initially normalized our high-resolution custom-collected screenshots and downscaled to $320 \times 352$ pixels, as pre-processing before model training. However, via an Ablation Study ($\S$\ref{section:size-res-ablation}) on our \textit{KhanAcademy} Interactable dataset, we observe that maximum-resolution inputs lead to significant performance gains with an acceptable sacrifice in inference speed. Our hyperparameter configuration is given in Appendix \ref{appendix:hparams}.

    \subsection{Subsystem 2: Screen Similarity Model}
    \label{section:screen-sim}
\begin{figure}[h!]
            \centering
            \includegraphics[width=\columnwidth]{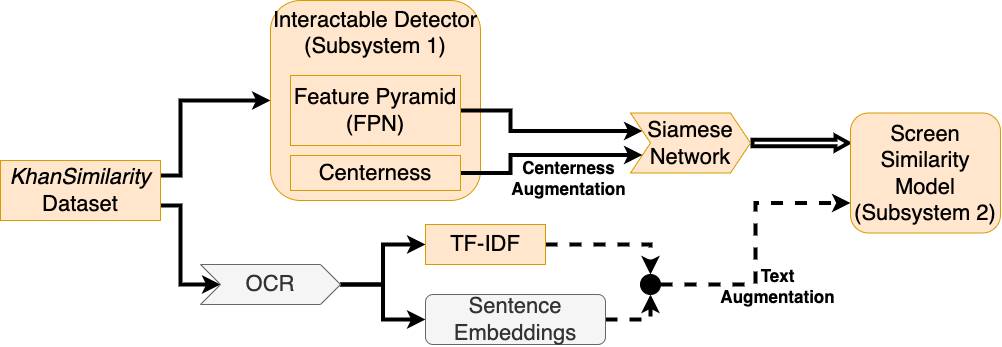}
            \caption{\textit{Screen Similarity Model} trained using a Siamese Network on pairs of our webpages labeled as ``same'' or ``different'' states. \textit{KhanSimilarity} denotes our dataset of \textit{KhanAcademy} screenshots for screen similarity.}
            \label{fig:screen-similarity-model}
        \end{figure}

        We develop our \textit{Screen Similarity} model to identify whether a pair of images belongs to the \textit{same} GUI state or \textit{different} states. This subsystem of the end-to-end \SystemName allows it to robustly identify state transitions, screen loading and matching interactables, using solely the GUI's pixel-appearance. The training pipeline for this model is shown in Figure~\ref{fig:screen-similarity-model}.

        Unlike prior work on state-similarity~\cite{webui,screenrel,neverending} that passed the raw GUI screenshots directly as inputs to a Siamese Network $G_W$, we utilize the interim Feature Pyramid (FPN) embedding features of our \textit{Interactable Detector} model as inputs to $G_W$. This allows us to (1) reuse the complex input-image features that have been already learnt by the ResNet backbone (C3-C5) and the further FPN (P3-P7), and also (2) introduce a \textit{further} learnable network $G_W$ that minimizes the \textit{Euclidean Distance} $d(\mathbf{E}_1,\mathbf{E}_2)$ between the final embeddings $\mathbf{E}_1,\mathbf{E}_2$ generated from the FPN input features.
                
        We hypothesize that the novel feature learning of our Siamese Network, enabled by utilizing FPN features as inputs, improves state-similarity detection over Wu~\etal's ResNet $G_W$. Via an Ablation Study ($\S$\ref{section:screensim-ablation}) on potential $G_W$ networks, we observe that even low-complexity networks are highly performant due to our efficient parameter reuse.

        \paragraph{FPN Input Feature Preprocessing}
            We observe that Euclidean Distance between the corresponding FPN feature maps P3-P7 of image pairs in our Interactable Detection datasets are \textit{correlated}; an image pair in \textit{KhanAcademy} has similar distance \textit{relative} to other pairs in the same feature map, across all P3-P7. Therefore, to maintain all FPN information \textit{w.l.o.g.}, we upsample all FPN maps to the size of the largest (P3) and concatenate them.

        \paragraph{Siamese Loss}
            Our Siamese Network $G_W$ implements the \textit{Contrastive Loss}~\cite{dimredux_siamese} on final embeddings $\mathbf{E}_1,\mathbf{E}_2$ produced from the input feature maps of two paired GUI screenshots:
            \begin{equation}
                \mathcal{L} = (1-Y)\cdot \max \{0, m_n - d(\mathbf{E}_1,\mathbf{E}_2) \} + Y \cdot \max \{ 0, d(\mathbf{E}_1,\mathbf{E}_2) - m_p \},
                \label{eq:contrastive}
            \end{equation}
            where $Y$ is the train-data \textit{label} ($Y=1$ for pair inputs being \textit{same} GUI state; $Y=0$ for pair inputs being \textit{different} states), and $m_p,m_n \in \mathbb{R}>0$ are \textit{margin} hyperparameters. In inference, two inputs are predicted to be the \textit{same} GUI state if $d(\mathbf{E}_1,\mathbf{E}_2) \leq (m_p + m_n)/2$.
            
        \paragraph{Siamese Architecture}            
            We considered multiple architectures for $G_W$. All contain an initial \textit{Linear} layer to learn the optimal combination of the input upsampled FPN feature maps P3-P7 for a GUI screenshot, as a weighted sum along the \textit{first} dimension of concatenation. This Linear layer is a fully-connected perceptron. We considered multiple ablations with ResNets~\cite{resnet,resnext} on top of the initial Linear layer ($\S$\ref{section:screensim-eval}). However, interestingly, the single Linear layer outperformed more complex ablations, by re-using image FPN feature maps as inputs. Meanwhile, overly complex ablations are prone to overfitting.

            A weakness of the current $G_W$ (Linear and ablations) is that the FPN input features are not \textit{linearly separable} and exhibit \textit{high overlap}. To solve this, we also extract \textit{Centerness} from the model alongside FPN P3-P7, and \textit{augment} the Siamese input. In the model, 5 Centerness feature maps emphasize the locations of predicted interactables in the corresponding P3-P7 FPN maps.
            
\begin{figure*}
                \centering
                \includegraphics[width=0.95\linewidth]{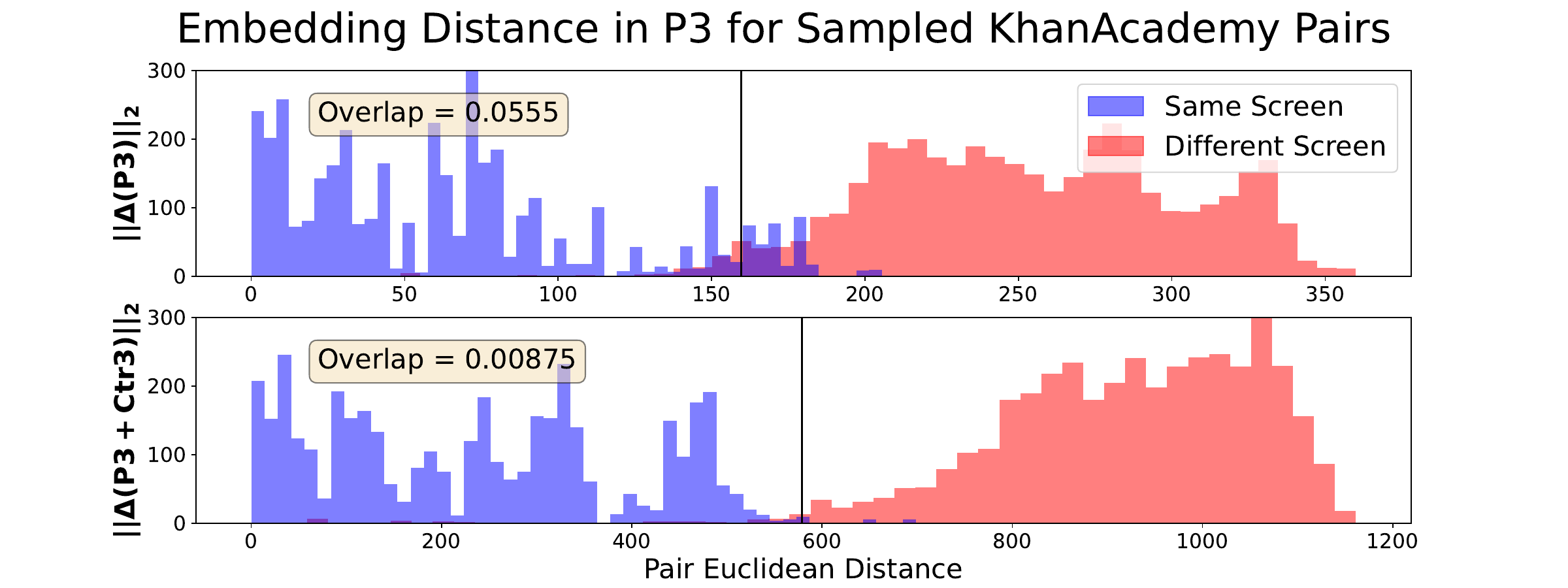}
                \caption{Pairs of screenshots of the same screen should be close to each other in the embedded space, in contrast to embeddings of different screens. The straight vertical line denotes the decision threshold. Here we visualize the Euclidean distances between 8000 training set pairs. On \textit{top}, the P3 embedding leaves a large number of pairs getting falsely mis-labeled. In contrast, P3 and with Centerness 3 (\textit{Ctr3} for short-hand) on the \textit{bottom} is more separable. }
                \label{fig:centerness_pair}
            \end{figure*}

            The power of Centerness-augmentation is shown in Figure \ref{fig:centerness_pair}, which plots the distribution of Euclidean Distance for (1) P3, and for (2) the addition of normalized\footnote{Normalization to zero mean and unity standard deviation.} P3 with the corresponding normalized Centerness feature map. This plot contains 8000 randomly-selected unique pairs of same- and different-state \textit{KhanAcademy} screens. Here, we measure data overlap in the \textit{input} to the Siamese $G_W$ as the proportion of same- and different-state points on the \textit{wrong} side of the decision threshold in the middle of the two distributions' centroids; we observe a six-fold increase in input-data separability with FPN-Centerness augmentation in Figure \ref{fig:centerness_pair}. Moreover, we provide the histograms comparing distributions (1) and (2) for all feature maps P3-P7 in Appendix \ref{appendix:centernessablation}.
            
            We observe that augmenting P3-P7 with Centerness strongly \textit{increases data separability} across all feature maps by a factor of $1.45$ to $6.34$. Further, we observe that Centerness augmentation skews the same- and different-state distributions away from each other. These characteristics enable the Siamese Network to learn a distinct distance-metric transformation with lower network complexity.
            
\begin{figure}[h!]
                \centering
                \includegraphics[width=0.95\columnwidth]{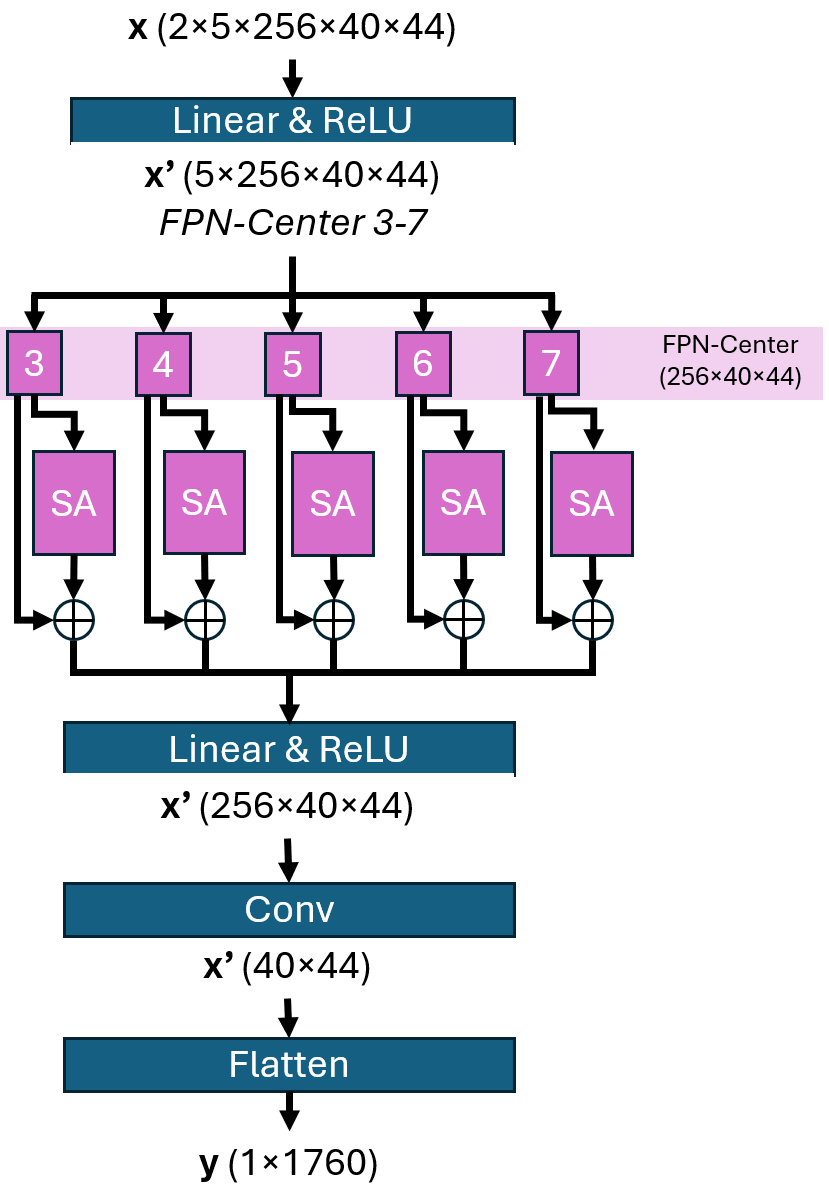}
                \caption{Centerness-augmented Siamese Network architecture, given an input as the upsampled concatenation of all FPN and Centerness feature maps. This is based on \textit{KhanAcademy} input dimensionality as an example.}
                \label{fig:screensim_fpncenter_arch}
            \end{figure}

            We now augment the prior $G_W$ by passing as input the \textit{concatenation} of upscaled FPN and Centerness features. Moreover, we augment the $G_W$ architecture, as presented in Figure \ref{fig:screensim_fpncenter_arch}. The first Linear layer implements a learnt combination of FPN-Centerness input features \textit{w.l.o.g.}. Then, for each combined FPN-Center feature map, a \textit{Self-Attention} (SA) layer learns point relationships over the entire feature map; to preserve the scale of the FPN-Centerness maps, we implement a feed-forward residual connection to the SA layers~\cite{sagan}. From \cite{sagan}, this is expressed as 
            
            \begin{flalign}
                SA(X) & := \mathrm{softmax}\left( \mathbf{q} \mathbf{k}^\top \right) \mathbf{v} \nonumber \\
                & = \mathrm{softmax} \left( \underbrace{\mathrm{conv}_{s=8}(X)}_{\mathbf{q}} \cdot \,{\underbrace{\mathrm{conv}_{s=8}(X)}_{\mathbf{k}}}^\top \right) \cdot \underbrace{\mathrm{conv}_{s=1}(X)}_{\mathbf{v}},
                \label{eq:selfattention}
            \end{flalign}
            where $s$ denotes a convolution's downsampling ratio relative to input dimensionality $|X|$. Finally, SA outputs are combined and convolved to learn cross-channel relationships across the feature-map channels.
            
            Overall, the Centerness-augmented Screen Similarity Model transforms the concatenated FPN-Centerness features of an image pair $i,j$ in a latent embedding space, such that final embeddings are nearby for same-state GUIs $i,j$.

        \paragraph{Text Augmentation}
            To aid in distinguishing text elements in text-dense webpages, we also explored adding a \textit{deterministic} augmentation external to $G_W$, which compares the OCR-extracted text content between the paired input images. To compare text, we implemented a custom-fitted Term Frequency-Inverse Document Frequency (TF-IDF) and sentence embeddings (Figure \ref{fig:screen-similarity-model}).

            On our Screen Similarity dataset for the \textit{KhanAcademy} website, with a downscaled resolution of $320\times 316$ pixels, we found that neither TF-IDF nor sentence embeddings significantly affect the performance of $G_W$. However, future work should explore the performance contribution of OCR-extracted features for higher-resolution inputs.

        \paragraph{Screen Similarity Summary}
            We follow a novel strategy of reusing previously-trained FPN and Centerness features from the \textit{Interactable Detector} model, as input to \textit{Screen Similarity} (hyperparameters in Appendix \ref{appendix:hparams}). This reduces model complexity, accelerates model training, and achieves superior performance competitive with computationally-complex GUI-based Siamese Networks~\cite{webui,screenrel} with a significantly smaller dataset ($284$ images on \textit{KhanAcademy}).

    \subsection{Subsystem 3: Action Matching}
    \label{section:actionmatch}

        Learning GUI visual semantics and robustly actuating the correct interactable selected by a user during a task demonstration are vital goals of an end-to-end GUI Automator. We define an end-to-end multi-step demonstration as a \textit{trace}; this consists of discrete GUI states transitioned due to user actions.
        
        However, it is not trivial to automatically locate the single correct interactable on an arbitrary GUI screen that a user demonstrated -- perhaps in a different device, OS, or application version. Indeed, existing end-to-end systems rely on heuristic exact-pixel pattern-matching~\cite{vasta} or repeat the user's demonstrated action on the exact coordinates logged during the demonstration~\cite{v2s,sugilite}. Instead, we introduce a novel Action Matching subsystem, which leverages our Interactable Detector and Screen Similarity subsystems for variation-robust and platform-independent interactable matching.

        The aim of the Action Matching subsystem is to identify the correct interactable $I_\mathrm{Replicate}$ during automatic replication of a user demonstration, which matches the interactable $I_\mathrm{Record}$ clicked\footnote{Or tapped in phone devices.} at on-screen position $(x,y)$ by the user during the demonstration recording.

        The input to the Action Matching subsystem is the screenshots $S_\mathrm{Record}, S_\mathrm{Replicate}$ in demonstration Recording and Replication, respectively, along with the user's click at $(x,y)$ during Recording. We assume that $S_\mathrm{Record}$ and $S_\mathrm{Replicate}$ are of the \textit{same} GUI state, and that there exists a solution $I_\mathrm{Target}$.

\begin{figure*}[h!]
            \centering
            \includegraphics[width=0.8\textwidth]{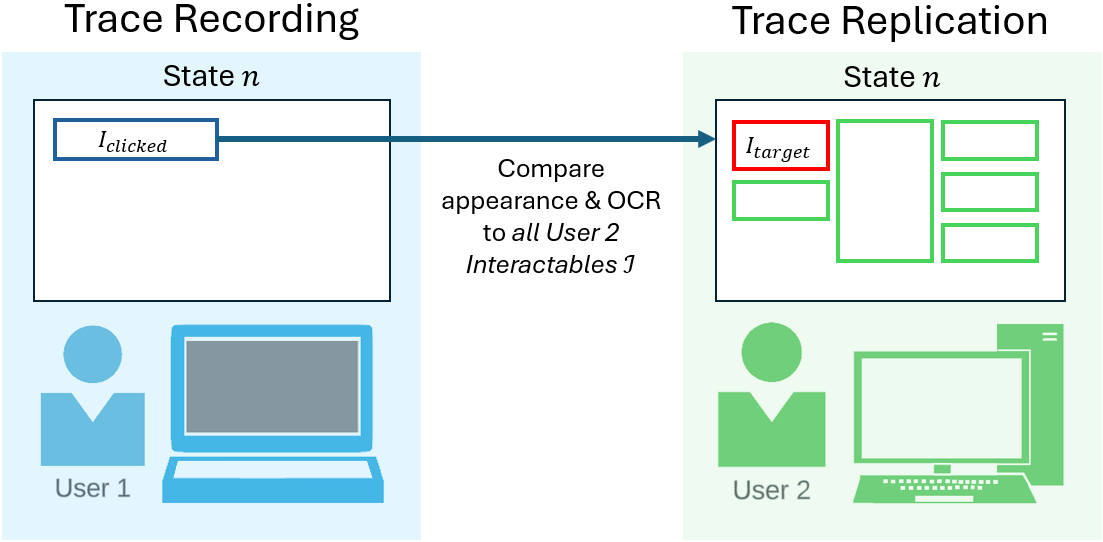}
            \caption{Action Matching the clicked interactable during Trace Recording to the best-match interactable during Trace Replication.}
            \label{fig:actionmatch}
        \end{figure*}

        In this process, we first identify all visible interactables $\mathcal{I}_\mathrm{Record}$, $\mathcal{I}_\mathrm{Replicate}$ in Recording and Replication phases, respectively. Then, we (1) determine which interactable was clicked during Recording using the recorded action position, and (2) compare its visual appearance and OCR-extracted text content to all $I_r\in \mathcal{I}_\mathrm{Replicate}$. In this comparison, we pass the FPN-Centerness feature maps of the pair $\{I_\mathrm{Record},I_r\}$ to our Screen Similarity model and measure the Euclidean distance of the output embeddings. Similarly, if both interactables in a pair have OCR text $T$, we convert both to sentence embeddings $E(T)$ and measure the Euclidean distance. The Replication interactable $I_r$ with smallest summed distance to $I_\mathrm{Record}$ is the target. This is formulated in Algorithm \ref{alg:actionmatch}, and visualized in Figure~\ref{fig:actionmatch}.
        
        Thereby, we robustly identify which \textit{target interactable} is the best match for Replication. Because Action Matching leverages our pure Computer Vision pipeline, we can identify the target even in translational on-screen shifts, minor appearance variations, background variations (\eg passive video playback), and changes in OS or device.

        \begin{algorithm*}[h!] \begin{algorithmic}[1]
            \STATE \textbf{Input:} Screenshots $S_\mathrm{Record}, S_\mathrm{Replicate}$, Action pixel-position $(x,y)$ from Recording
                    
            \STATE $\mathcal{I}_\mathrm{Record} \longleftarrow \mathrm{InteractableDetector}(S_\mathrm{Record})$ %

            \STATE $I_\mathrm{Record} \longleftarrow I\in\mathcal{I}_\mathrm{Record} \,\, | \,\, (x,y)\in I$

            \STATE $\mathcal{I}_\mathrm{Replicate} \longleftarrow \mathrm{InteractableDetector}(S_\mathrm{Replicate})$

            \STATE            $I_\mathrm{Target} = \underset{I_r \in \mathcal{I}_\mathrm{Replicate}}{\mathrm{argmin}} \left( \mathrm{ScreenSimilarity}(I_\mathrm{Record}, I_r) + \big\lVert \left|E(T_\mathrm{Record}) - E(T_r) \right| \big\rVert_2 \right)$
                    
            \STATE \textbf{Return:} Output the target interactable $I_\mathrm{Target}$
        \end{algorithmic} \caption{Action Matching} \label{alg:actionmatch} \end{algorithm*}

    \subsection{Voice Navigation}
    
        A pertinent use case for the \textit{Interactable Detector} is GUI Navigation using voice commands. The \textit{Interactable Detector} can highlight interactables on the screen, and the user can then specify which interactable to press using voice as input. Subsequently, the \SystemName completes the action and waits for the next prompt from the user. This empowers the user by allowing hands-free navigation of accessibility-lacking applications and websites.
                
        We implemented a proof-of-concept application using OpenAI's \textit{Whisper}~\cite{whisper} speech-to-text model. After starting the application, the system listens for one of three commands: (1) \texttt{show}, (2) \texttt{click <number>}, and (3) \texttt{close}. The \texttt{show} command displays bounding boxes for detected interactables, and enumerates them. After the user chooses a specific numbered interactable using \texttt{click <number>}, the system performs a left mouse-click in the center of the specified interactable's bounding box (as predicted by the \textit{Interactable Detector}). The user can repeat this process indefinitely, and may call \texttt{close} command to end the session.
                
        Hands-free GUI navigation places the \textit{Interactable Detector} in an inherently-practical context, which allows us to test its real-life usability. Indeed, besides the mAP metric, a \textit{usable} \textit{Interactable Detector} should provide the most comprehensive list of actionable interactables. Hence, this proposed application doubles as a benchmarking model against baseline models like Wu~\etal's detector~\cite{webui}. Beyond the mAP performance reported in $\S$\ref{section:interactabledetector}, experiments in Sec.~\ref{section:gui-results-nav} seek to answer the question: ``Given an image and a target interactable\footnote{Indicated by user voice command.}, can a chosen model identify a valid screen location to interact with the target?''

    \subsection{Trace Life-Cycle}
        We propose a framework for general representation and replication of the set of user actions in the computer system. The center of it is the \textit{Trace} - set of states and actions connecting them representing the human-like navigation of the GUI. We implement the framework for mouse left-click actions and simple keyboard inputs, though it can be extended to accommodate more complex keyboard shortcuts, mouse right-clicks, drag-and-drops, and other actions. This general framework proposes how our \SystemName subsystems ($\S$\ref{section:interactabledetector}-$\S$\ref{section:actionmatch}) should interconnect to \textbf{record} a user demonstration, \textbf{store} and \textbf{replicate} it on same or different system. Every step of the proposed structure minimizes platform dependency in some sort of way.
        
        \begin{figure}[h!]
            \centering
            \includegraphics[width=\linewidth]{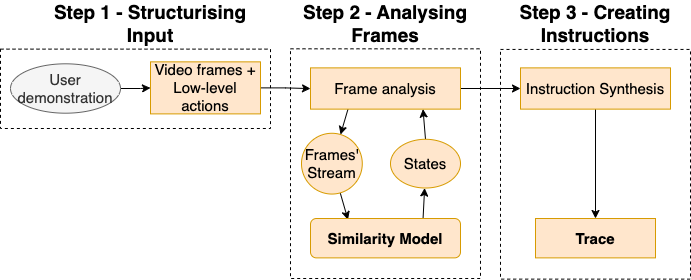}
            \caption{Process for a user to record a Trace as a video of GUI frames and timestamped low-level actions.}
            \label{fig:trace-creation-flow}
        \end{figure}
        
        Our Trace Recording process is outlined below and visualized in Figure \ref{fig:trace-creation-flow}:
        \begin{itemize}
            \item[Step 1:] Stores the frames and user actions in a format common for Unix (\eg MacOS, Linux), Windows, and Android devices;
            \item[Step 2:] Uses \textit{State Similarity} to segment the continuous video in Trace Creation into distinct same-state blocks, and stores a representative screenshot for each block; and
            \item[Step 3:] Creates a small collection of files representing the trace - a set of actions and accompanying screenshot for each action.
        \end{itemize}
        
        \begin{figure}[h!]
            \centering
            \includegraphics[width=\linewidth]{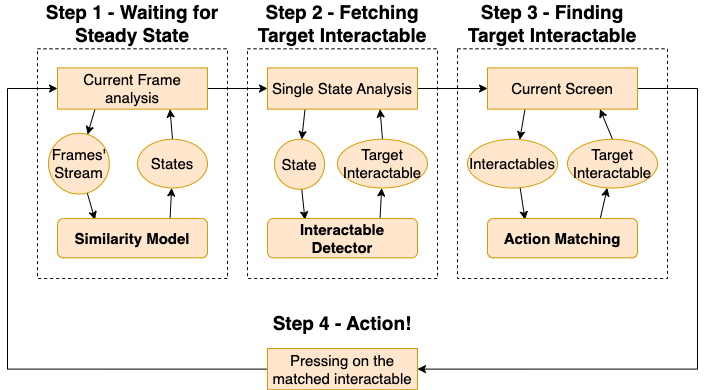}
            \caption{Process for our \SystemName to automatically reproduce the recorded Trace in a variety of devices and OSs.}
            \label{fig:trace-reproduction-flow}
        \end{figure}
        
        During the Trace Replication phase visualized in Figure \ref{fig:trace-reproduction-flow}, our \SystemName acts as follows:
        \begin{itemize}
            \item[Step 1:] Waits for a steady state;
            \item[Step 2:] From the recorded trace, identifies which interactable the user pressed;
            \item[Step 3:] From the current state finds the interactable matching the one from Step 2 using Action Matching; and
            \item[Step 4:] Actuates the interactable and resets to Step 1 for the next action.
        \end{itemize}

        We populate this framework with our own \textit{Interactable Detector}, \textit{State Similarity} and \textit{Action-Matching} models. Even though, in the future these subsystems could be replaced by better-performing architectures and the framework itself incorporated into Never-Ending Learning paradigms, we believe that this way of structuring the stream of data is a key contribution to the field of GUI-Automators. It enables stateful representation of the GUI environment, common -- but still flexible -- structure for interfacing the states, and enables cross-platform functionality.

        We continue by discussing our implementation of the steps in the framework.
    
        \paragraph{Trace Recording}
            When recording a trace demonstrated by a user, the \SystemName records (1) GUI screenshots at a sufficient frame-rate and (2) the user's low-level actions. Afterwards, the gathered data is processed and the resultant trace recording is structured as JSON. Every state-action pair contains the state-defining frame screenshot and the high-level action which was executed. In a trace, there are $N$ GUI states and $N-1$ actions.
        
            \begin{figure*}[h!]
                \centering
                \includegraphics[width=\textwidth]{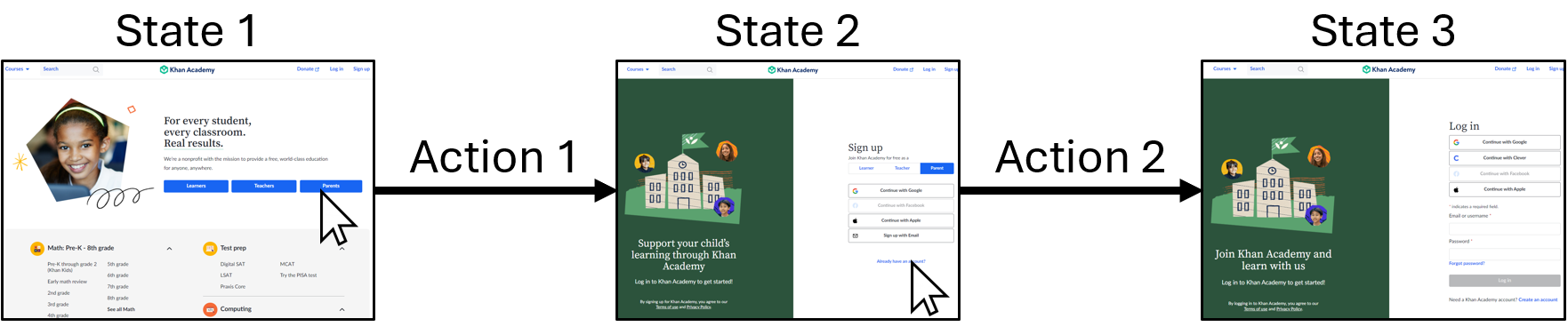}
                \caption{Example trace in \textit{KhanAcademy}, where user actions cause a GUI state change.}
                \label{fig:tracegraph}
            \end{figure*}

\begin{figure}[h!]
                \centering
                \includegraphics[width=0.95\columnwidth]{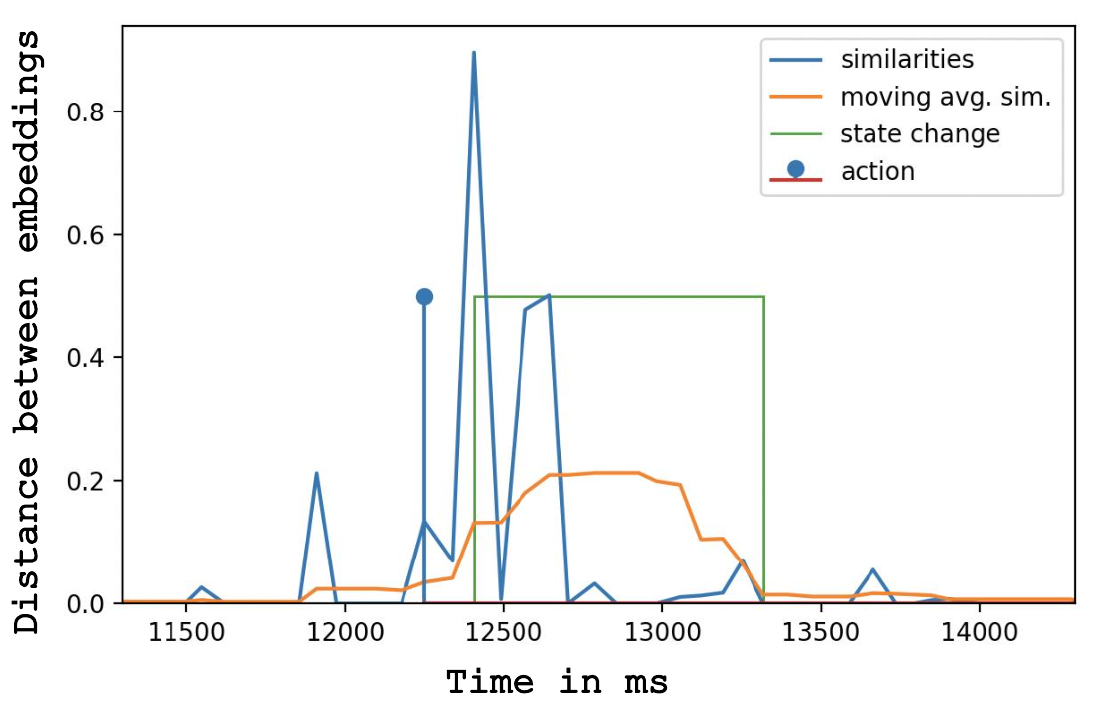}
                \caption{Time series data showing similarity of consecutive frames, along with associated data. $0$ indicates similar frames, steady state, and no action. A distance threshold of $0.35$ delineates state transitions. }
                \label{fig:single-state-change-plot}
            \end{figure}

            When processing the stream of images from a trace recording, the \SystemName finds the similarity between all consecutive pairs using \textit{Screen Similarity} (Figure \ref{fig:single-state-change-plot}). Commonly, while the screen is loading during a state transition, it may change repeatedly for multiple frames before stabilizing on the next state. Thus, we use a moving average window on the screen similarity curve to judge when the GUI state changes and stabilizes. The blocks of consecutive frames with steady state form the distinct states reached by the user during the demonstration. Finally, for each multiple-frame block of the same stable state, we represent it with the first frame after the user's action and before the resultant state change begins. This discretized representation significantly reduces clutter and enables downstream processing tasks.

            To our knowledge, this ML-based method of identifying steady and changing state in a GUI environment is superior to prior methods. Our method is simultaneously sensitive to user actions during the trace recording while also neglecting GUI changes not caused by a user action (\eg computer clock ticking, pixels updating on a currently-playing video, loading screen between websites, etc.). We thus build a robust trace representation, without relying on any platform-specific APIs.
            
        \paragraph{Trace Replication}
            The \SystemName is capable of trace replication on any machine, using only GUI screenshots recorded as frames on the machine. To successfully replicate a user-recorded trace, the \SystemName must identify:
        
            \begin{enumerate}
                \item When is the state stable on the machine?
                \item What is the \textit{target interactable} to click on at the current stable state?
            \end{enumerate}
                
\begin{figure*}[h!]
                \centering
                \includegraphics[width=\textwidth]{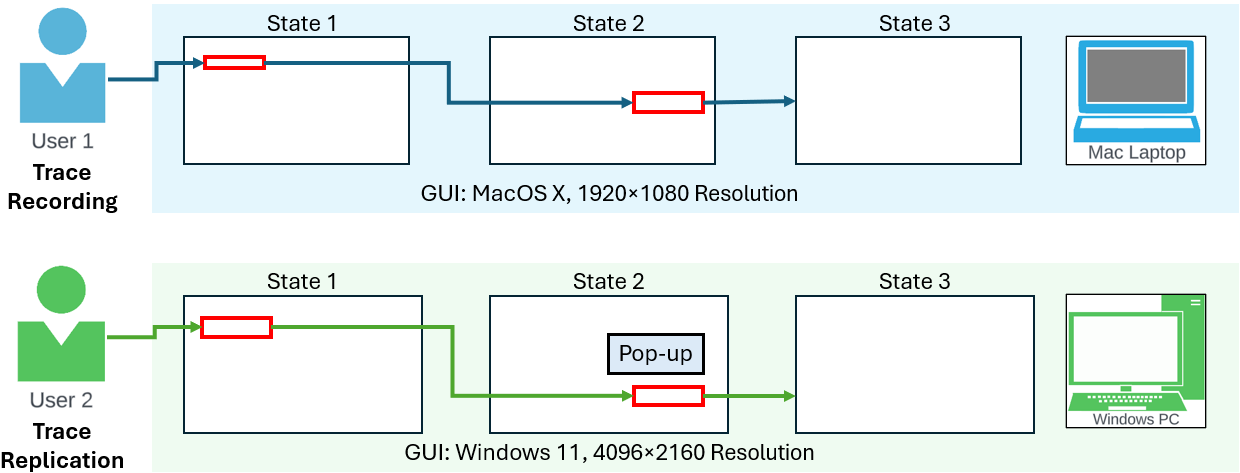}
                \caption{Example of Trace Replication on a different machine than Trace Recording, with visual GUI changes due to platform changes and pop-ups.}
                \label{fig:traceenv}
            \end{figure*}

            Point (1) is necessary to progress to Point (2), namely clicking the same (\textit{target}) interactable on the same state as the recorded trace. As in Trace Creation, we achieve Point (1) by a consecutive-frame real-time analysis with Screen Similarity. However, Point (2) is not trivial. This is because a multitude of visual changes can occur in the GUI environment during Trace Recording and Trace Replication, as shown in Figure \ref{fig:traceenv}. For instance, updates to GUI applications (\eg website redesign) or shifts in device and OS cause appearance changes in the target interactable. Meanwhile, pop-up boxes or changes in screen pixel resolution cause translational changes in the target interactable's location on the screen. Therefore, past works~\cite{vasta,v2s} that repeat an action at the same pixel position as in Trace Recording are not robust in Trace Replication.
                
\section{Experiments}

We now evaluate our \SystemName in terms of our key contributions: (1) performance of Interactable detection and Screen Similarity on our high-quality data collected with our novel open-sourced framework; (2) real-life voice-controlled GUI navigation; and (3) robust platform-independent Trace Replication for synthetically-generated traces. All of our ML subsystems were trained in a High-Performance Computing GPU cluster using data-parallelized training across 8 GPUs (NVIDIA V100), with \texttt{pytorch} multi-GPU training optimizations.

This section presents experiments and ablations as follows, addressing each of our key contributions:
\begin{enumerate}[i]
    \item Evaluation of the Interactable Detector separately trained on our data for each of our target GUIs. We explore the effects of dataset size and input resolution in an ablation ($\S$\ref{section:interactable-eval}).
    \item Evaluation of the Centerness-augmented Screen Similarity model separately trained on \textit{KhanAcademy} and \textit{Spotify}. We also explore various FPN-only architectures in an ablation, showing the surprising strength of the Linear $G_W$ ($\S$\ref{section:screensim-eval}).
    \item Hands-free voice-navigation of the \textit{KhanAcademy} GUI using our Interactable Detector or other works' detectors~\cite{webui,vins,easyocr} to recommend available interactables. We show that our Interactable Detector is deployable for real-time navigation ($\S$\ref{section:gui-results-nav}).
    \item Cross-platform Trace Replication on real traces for the \textit{KhanAcademy GUI}. We show that our end-to-end \SystemName can robustly replicate user traces ($\S$\ref{section:gui-results-trace}).
\end{enumerate}

\subsection{Evaluation of Interactable Detection on Collected GUI Data}
\label{section:interactable-eval}

We demonstrate that we can achieve excellent detection performance for interactables across multiple devices, Operating Systems, and application types, using a minimal dataset of only a few hundred to a few thousand samples for each target GUI. Perhaps expected, fine-tuning from generic multi-GUI datasets~\cite{webui,vins} does not guarantee a performance benefit. Indeed, this is also shown by the plateau in detection performance on the multi-GUI never-ending learning by Wu~\etal~\cite{neverending}. 

While sometimes previously considered bad experimental hygiene, optimizing for test-scenarios is now a respected pathway for some ML systems~\cite{liang2024comprehensive}. We argue that this shift is justified to maximize the user's experience. Consequently, honing on a smaller but focused dataset is attractive too, while broad but ``shallow'' datasets across a plethora of GUIs, the norm in current GUI Automation systems~\cite{webui,screenrel,spotlight,neverending,vins,rico}, is useful for reasons of generalization.

Specifically, we train our Interactable Detector model separately on the training set (80\%) of each target GUI in $\S$\ref{section:dataacq}. For each target GUI, we train the detector both (1) from scratch and (2) fine-tuned from a checkpoint pre-trained on 7k noise-cleaned websites gathered by Wu~\etal~\cite{webui} or 4.8k Android phones in VINS~\cite{vins}. Further, we compare this with training solely on Wu~\etal's 7k websites or VINS, without fine-tuning on our data. The input screenshots are maximum-resolution without down-sampling. All models were trained with early-stopping for a minimum of 150 epochs, with the same hyperparameters tuned for maximum mAP on the validation set of \textit{KhanAcademy}. See Appendix~\ref{appendix:hparams} for the hyperparameter configuration.

Finally, we evaluate model performance on the held-out validation (10\%) and testing (10\%) sets using mAP at IoU confidence threshold $0.5$. We additionally report test-time inference speed in frames-per-second (FPS), to examine the real-time deployment of our end-to-end \SystemName. This is shown in Table \ref{tab:fcos-interactable-results}. Indeed, scratch-trained performance largely tops fine-tuned performance, with the varying performance gap dependent on how closely the target GUI's appearance distribution matches the pre-training dataset. Here, we observe the limited generalization of the pre-trained \textit{web7k} and \textit{VINS} to our target-GUI datasets, shown by the very poor zero-shot performance of the non-finetuned \textit{web7k-only} and \textit{VINS-only} models on our target GUIs.

\begin{table*}[t]
    \centering
    \caption{Training, Validation and Test performance of our Interactable Detector across multiple custom-gathered target GUI websites and applications in computers and Android phones. \textit{Scratch-<GUI>} denotes model training only on the dataset of our target GUI; \textit{web7k-<GUI>} denotes fine-tuning on the target GUI with \textit{web7k} pre-training (similar for \textit{VINS}); \textit{web7k-only} denotes training only on \textit{web7k} (similar for \textit{VINS}). \textsuperscript{\textdagger}\textit{Phone} is short-hand for the \textit{Telephone \& Contacts} GUI app.}
    \begin{tabular}{llcccc}
        \toprule
        Target GUI & Model & Train mAP@$0.5$ & Val mAP@$0.5$ & Test mAP@$0.5$ & Test FPS \\
        \midrule

        \multirow{3}{*}{\textit{KhanAcademy}} & \textit{Scratch-Khan} & $\mathbf{0.9663}$ & $\mathbf{0.9583}$ & $\mathbf{0.9710}$ & $7.62$ \\
         & \textit{web7k-Khan} & $0.6592$ & $0.6481$ & $0.6506$ & $18.79$ \\
         & \textit{web7k-only} & $0.0032$ & $0.0032$ & $0.0029$ & $19.33$ \\
        \midrule

        \multirow{3}{*}{\textit{Wolfram-Alpha}} & \textit{Scratch-Wolfram} & $\mathbf{0.7865}$ & $\mathbf{0.7960}$ & $\mathbf{0.7805}$ & $4.23$ \\
         & \textit{web7k-Wolfram} & $0.7702$ & $0.7684$ & $0.7406$ & $14.33$ \\
         & \textit{web7k-only} & $2.5\times 10^{-4}$ & $3.1\times 10^{-4}$ & $1.9\times 10^{-4}$ & $14.81$ \\
        \midrule

        \multirow{3}{*}{\textit{Wikipedia}} & \textit{Scratch-Wikipedia} & $\mathbf{0.7254}$ & $\mathbf{0.7048}$ & $\mathbf{0.7317}$ & $4.03$ \\
         & \textit{web7k-Wikipedia} & $0.5031$ & $0.4837$ & $0.5043$ & $11.45$ \\
         & \textit{web7k-only} & $2.7\times 10^{-6}$ & $6.0\times 10^{-6}$ & $2.8\times 10^{-5}$ & $11.05$ \\
        \midrule

        \multirow{3}{*}{\textit{Spotify}} & \textit{Scratch-Spotify} & $\mathbf{0.9563}$ & $\mathbf{0.9489}$ & $\mathbf{0.9356}$ & $11.56$ \\
         & \textit{VINS-Spotify} & $0.9451$ & $0.9325$ & $0.9243$ & $19.10$ \\
         & \textit{web7k-Spotify} & $0.9283$ & $0.9131$ & $0.9114$ & $22.27$ \\
         & \textit{VINS-only} & $0.0126$ & $0.0117$ & $0.0185$ & $22.3$ \\
        \midrule

        \multirow{3}{*}{\textit{Telephone \& Contacts}} & \textit{Scratch-Phone}\textsuperscript{\textdagger} & ${0.9604}$ & ${0.9573}$ & $\mathbf{0.9669}$ & $12.46$ \\
         & \textit{VINS-Phone}\textsuperscript{\textdagger} & $\mathbf{0.9670}$ & $\mathbf{0.9599}$ & $0.9646$ & $22.60$ \\
         & \textit{web7k-Phone}\textsuperscript{\textdagger} & $0.9476$ & $0.9500$ & $0.9598$ & $22.47$ \\
         & \textit{VINS-only} & $0.0175$ & $0.0137$ & $0.0225$ & $20.77$ \\
        
        \bottomrule
    \end{tabular}
    \label{tab:fcos-interactable-results}
\end{table*}

\subsubsection{Ablation: Data Resolution and Dataset Size for Detection}
\label{section:size-res-ablation}

We assert that we can hone on a singular GUI environment and obtain strong detection performance with a small-scale high-quality training dataset. Here, we explore the rate of deterioration in detection performance when we (1) down-sample the input screenshots to lower resolutions or (2) reduce the training dataset size. At the cost of reduced detection performance, these choices decrease model training time and improve the inference speed (FPS).

Firstly, for the Data Resolution ablation, we explore down-sampling our \textit{KhanAcademy} dataset, with resolution $1249\times 1234$. We then down-sample the \textit{KhanAcademy} screenshots by proportionately down-scaling such that the smaller dimension is $640$ and $320$ pixels. Secondly, for the Dataset Size ablation, we randomly sub-sample our \textit{KhanAcademy} Interactable training dataset, and average performance metrics over multiple training sessions to account for random variations.

Table \ref{tab:resablation} jointly displays the two ablations. This shows that at lower input resolutions, the model trains faster and boasts a significant FPS improvement; however, the trade-off is detection performance. Real-time systems should strike a fine balance between sufficient inference speed at maximum detection accuracy. Further, we interestingly observe a gradual performance degradation when training on half of the full training set and evaluating on the \textit{same} validation and test sets. Hence, in situations of data scarcity, it is still feasible to attain sufficient detection performance.

\begin{table*}
    \centering
    \caption{Ablations for Training Dataset Size and Input Resolution for \textit{KhanAcademy} Interactable Detection. We sweep dataset sizes of 100 until full training-set (2272 screenshots); we sweep input resolutions across $1234$ (full-resolution), $640$ and $320$ pixels.}
    \begin{tabular}{cccccccc}
         \toprule
         \multirow{2}{*}{Train Size} & \multirow{2}{*}{Resolution} & \multirow{2}{*}{Runs} & \multirow{2}{*}{Training (hrs)} & \multirow{2}{*}{Test FPS} & \multicolumn{3}{c}{mAP$@0.5\ [\mu\pm\sigma]$} \\
         \cmidrule{6-8}
         & & & & & Train & Val & Test \\
         \midrule
         \multirow{3}{*}{100}  & 320 & 5   & 0.22      & 17.7 $\pm$ 1.1 & 0.372 $\pm$ 0.102 & 0.349 $\pm$ 0.088 & 0.338 $\pm$ 0.085 \\
     & 640 & 5   & 0.26      & 13.8 $\pm$ 0.9 & 0.542 $\pm$ 0.030 & 0.532 $\pm$ 0.024 & 0.526 $\pm$ 0.026 \\
     & 1234 & 10 & 0.40      & 7.9 $\pm$ 0.05 & 0.701 $\pm$ 0.082 & 0.706 $\pm$ 0.070 & 0.703 $\pm$ 0.076 \\ \midrule
\multirow{3}{*}{500}  & 320 & 5   & 0.49      & 17.2 $\pm$ 1.1 & 0.321 $\pm$ 0.028 & 0.313 $\pm$ 0.027 & 0.301 $\pm$ 0.027 \\
     & 640 & 5   & 0.58      & 14.2 $\pm$ 0.8 & 0.375 $\pm$ 0.144 & 0.375 $\pm$ 0.144 & 0.367 $\pm$ 0.146 \\
     & 1234 & 5  & 1.95      & 7.5 $\pm$ 0.04 & 0.719 $\pm$ 0.342 & 0.713 $\pm$ 0.339 & 0.719 $\pm$ 0.343 \\ \midrule
\multirow{3}{*}{1000} & 320 & 5   & 0.90      & 17.5 $\pm$ 0.8 & 0.449 $\pm$ 0.006 & 0.440 $\pm$ 0.006 & 0.431 $\pm$ 0.006 \\
     & 640 & 5   & 1.07      & 15.5 $\pm$ 0.1 & 0.666 $\pm$ 0.014 & 0.656 $\pm$ 0.017 & 0.661 $\pm$ 0.015 \\
     & 1234 & 5  & 3.87      & 7.4 $\pm$ 0.02 & 0.943 $\pm$ 0.005 & 0.935 $\pm$ 0.003 & 0.948 $\pm$ 0.004 \\ \midrule
\multirow{3}{*}{2272} & 320 & 4   & 1.77      & 19.6 $\pm$ 0.3 & 0.608 $\pm$ 0.015 & 0.601 $\pm$ 0.015 & 0.592 $\pm$ 0.018 \\
     & 640 & 4   & 2.08      & 15.9 $\pm$ 0.05 & 0.838 $\pm$ 0.016 & 0.828 $\pm$ 0.015 & 0.831 $\pm$ 0.016 \\
     & 1234 & 3  & 7.63      & 7.6 $\pm$ 0.01 & \textbf{0.964} $\pm$\textbf{ 0.001} & \textbf{0.956} $\pm$ \textbf{0.002} & \textbf{0.970} $\pm$ \textbf{0.001} \\ \bottomrule
    \end{tabular}
    \label{tab:resablation}
\end{table*}

\subsection{Evaluation of Screen Similarity on Collected GUI Data}
\label{section:screensim-eval}

We achieve \textit{state-of-the-art} results in state-similarity classification, across multiple devices and Operating Systems. Significantly, via our parameter-efficient architecture and Centerness-augmentation, we compete with and outperform existing Screen Similarity works~\cite{webui,screenrel,neverending} with a dataset smaller by several orders of magnitude.

We train our Screen Similarity model by randomly sampling batches of same-state or different-state pairs from our datasets. Again following our principle of honing in on singular GUI applications, we train a separate Screen Similarity model on \textit{KhanAcademy} (computer) and \textit{Spotify} (Android phone) to demonstrate cross-platform applicability. Every 10 training epochs, we validate on 50 randomly-selected pairs from the Validation set, and monitor the classification accuracy and F1 metrics. We stop training via early-stopping that monitors the F1 metric. Our hyperparameter configuration for both \textit{KhanAcademy} and \textit{Spotify} training is given in Appendix \ref{appendix:hparams}. Table \ref{tab:screensim_results} shows the final accuracy and F1 score for the Centerness-augmented $G_W$ on web-based \textit{KhanAcademy} and Android \textit{Spotify}.

Our performance in Table \ref{tab:screensim_results} is competitive with Wu~\etal's ResNet Siamese Network trained on 70k and 350k examples~\cite{webui}. Additionally, we outperform the Siamese \textit{ResNet} (F1 $0.69$) and Siamese \textit{Faster R-CNN} (F1 $0.83$) attained by Feiz~\etal~\cite{screenrel} on approximately 33k crowd-sourced smartphone screenshots. Similarly, we outperform the never-ending Siamese architecture (F1 $0.663$) of Wu \etal~\cite{neverending}, which was pre-trained on over 1250 multi-GUI screenshots (800k pairs) and fine-tuned during a live traversal on 140 screenshots (10k pairs). Meanwhile, our \textit{KhanAcademy} and \textit{Spotify} screen similarity datasets consist of $378$ and $451$ screenshots, respectively, which are 1-2 orders of magnitude smaller than the aforementioned works. Further, unlike~\cite{screenrel,neverending}, we release our \textit{KhanAcademy} dataset publicly for benchmarking.

Overall, our parameter- and feature-efficient $G_W$ with FPN-Centerness inputs is a strong choice for single-GUI similarity detection. Moreover, even a small dataset of approximately 400 screenshots yields 80k unique pair combinations. Thus, we assert that a large raw-screenshot dataset such as~\cite{webui,screenrel,neverending} is not necessary for distribution-representative training and high similarity detection performance.

\begin{table}[h!]
    \centering
    \caption{Validation and Test performance of the Centerness-augmented Screen Similarity model with the $G_W$ architecture consisting of Linear, Self-Attention and Convolution blocks.}
    \begin{tabular}{l c c c}
        \toprule
        Target GUI & Evaluation Set & Accuracy & F1 \\
        \midrule
        \multirow{2}{*}{\textit{KhanAcademy}} & Validation & $0.9605$ & $0.8276$ \\
        & Test & $0.9916$  & $0.9123$ \\
        \midrule
        \multirow{2}{*}{\textit{Spotify}} & Validation & $0.9858$ & $0.8649$ \\
        & Test & $0.9628$  & $0.7778$ \\
        \bottomrule
    \end{tabular}
    \label{tab:screensim_results}
\end{table}

\subsubsection{Architectural Ablations for FPN-Only $G_W$}
\label{section:screensim-ablation}

We briefly outline our architectural ablations for the Siamese Network $G_W$ with FPN-only input, tested with our \textit{KhanAcademy} screen similarity dataset. This justifies our efficient re-use of FPN parameters as it allows for a computationally-simple and surprisingly powerful $G_W$ Linear network. Beyond the simple Linear architecture, we also consider adding ResNet~\cite{resnet} or ResNeXt~\cite{resnext} layers, with or without instance-normalization and ReLU intermediary activations. Table \ref{tab:screensim_ablation} shows that the Linear architecture achieves the best F1.

\begin{table}[h!]
    \centering
    \caption{Validation and Test performance of non-Centerness ablations for $G_W$ on our \textit{KhanAcademy} data.}
    \begin{tabular}{l c c}
    \toprule
        Model $G_W$ & F1 \\
        \midrule
        \textit{ResNeXt} & $0.6489$ \\
        \textit{ResNeXt no-ReLU} & $0.5650$ \\
        \textit{ResNeXt no-Instance-Norm} & $0.6791$ \\
        \textit{ResNet} & $0.5072$ \\
        \textit{Linear} & $\mathbf{0.7516}$ \\
        \bottomrule
    \end{tabular}
    \label{tab:screensim_ablation}
\end{table}

\begin{figure}[h!]
    \centering
    \begin{subfigure}
        \centering
        \includegraphics[width=0.375\textwidth]{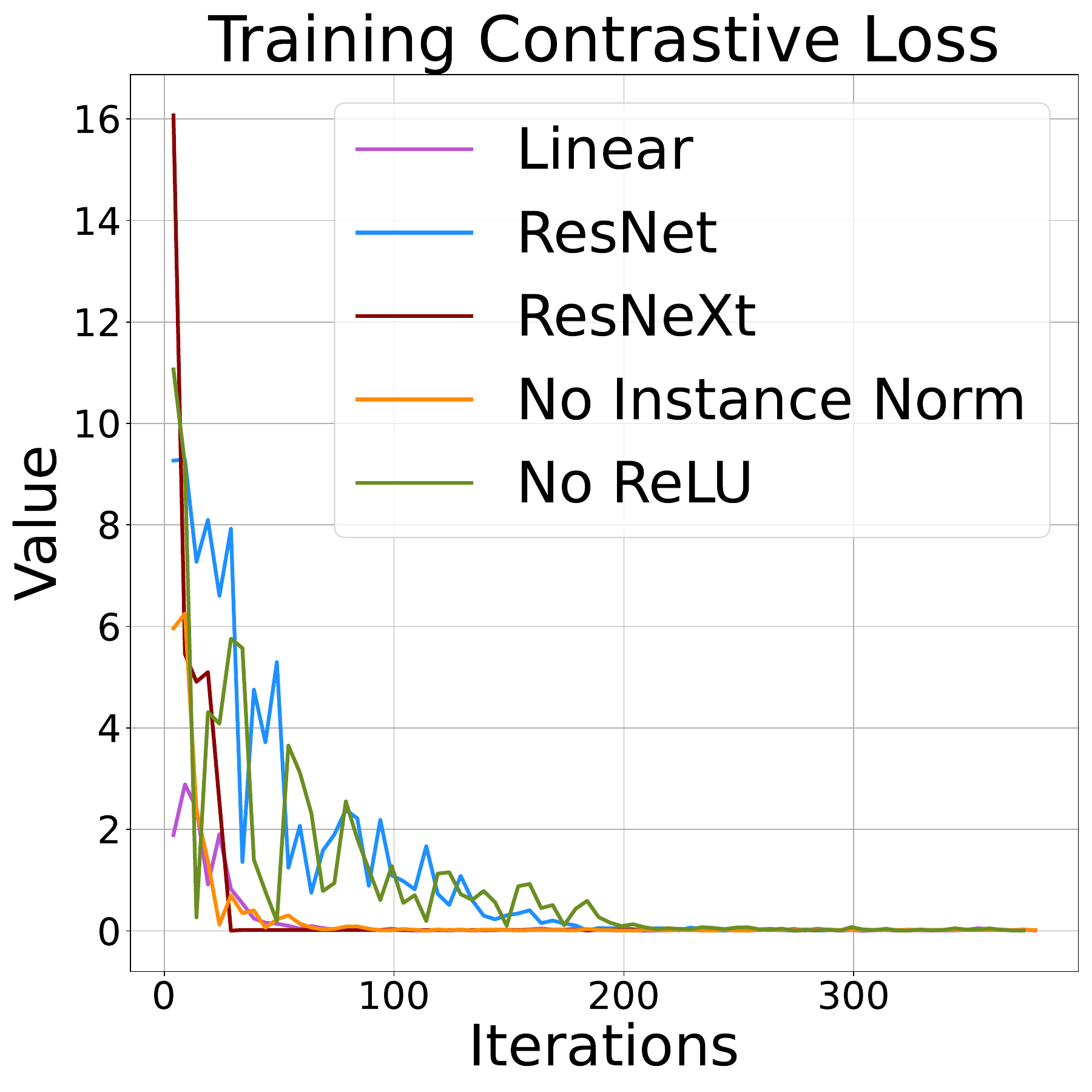}
    \end{subfigure}%
    \begin{subfigure}
        \centering
        \includegraphics[width=0.375\textwidth]{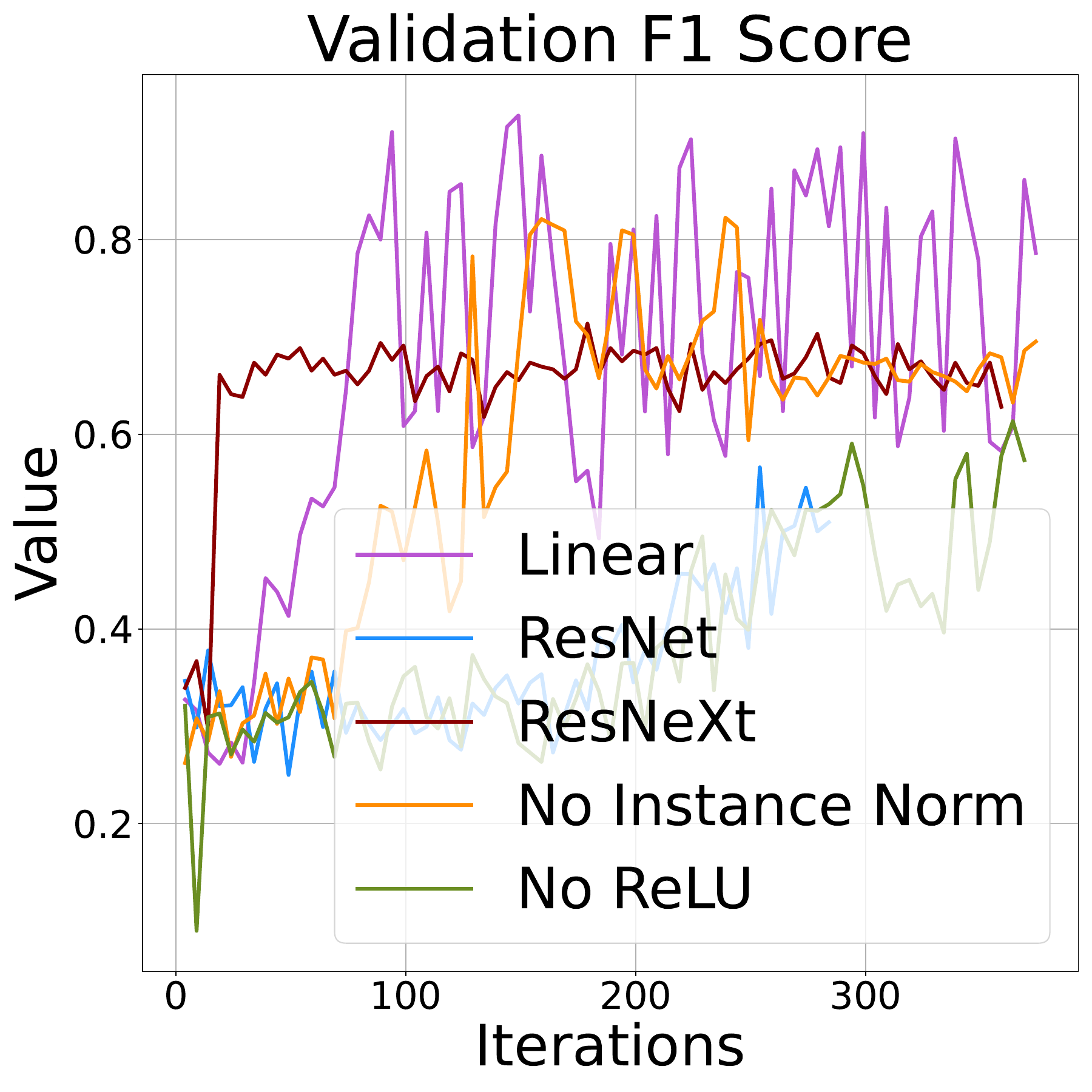}
    \end{subfigure}
    \caption{A linear $G_W$ is surprisingly effective. We found it out-performed more complex models, illustrated here with better training convergence and higher F1 validation scores.}
    \label{fig:screensim_fpn_train}
\end{figure}

\subsection{Automated GUI Navigation}
\label{section:gui-results-nav}

We show that GUI navigation using our Interactable Detector is successful in real-life GUI traversals within \textit{KhanAcademy}, more so than navigation using other existing detectors, though OCR-based methods showed strengths in text-heavy interfaces.

\begin{figure}[h!]
    \centering
    \includegraphics[trim=0 50 250 0,clip,width=\columnwidth]{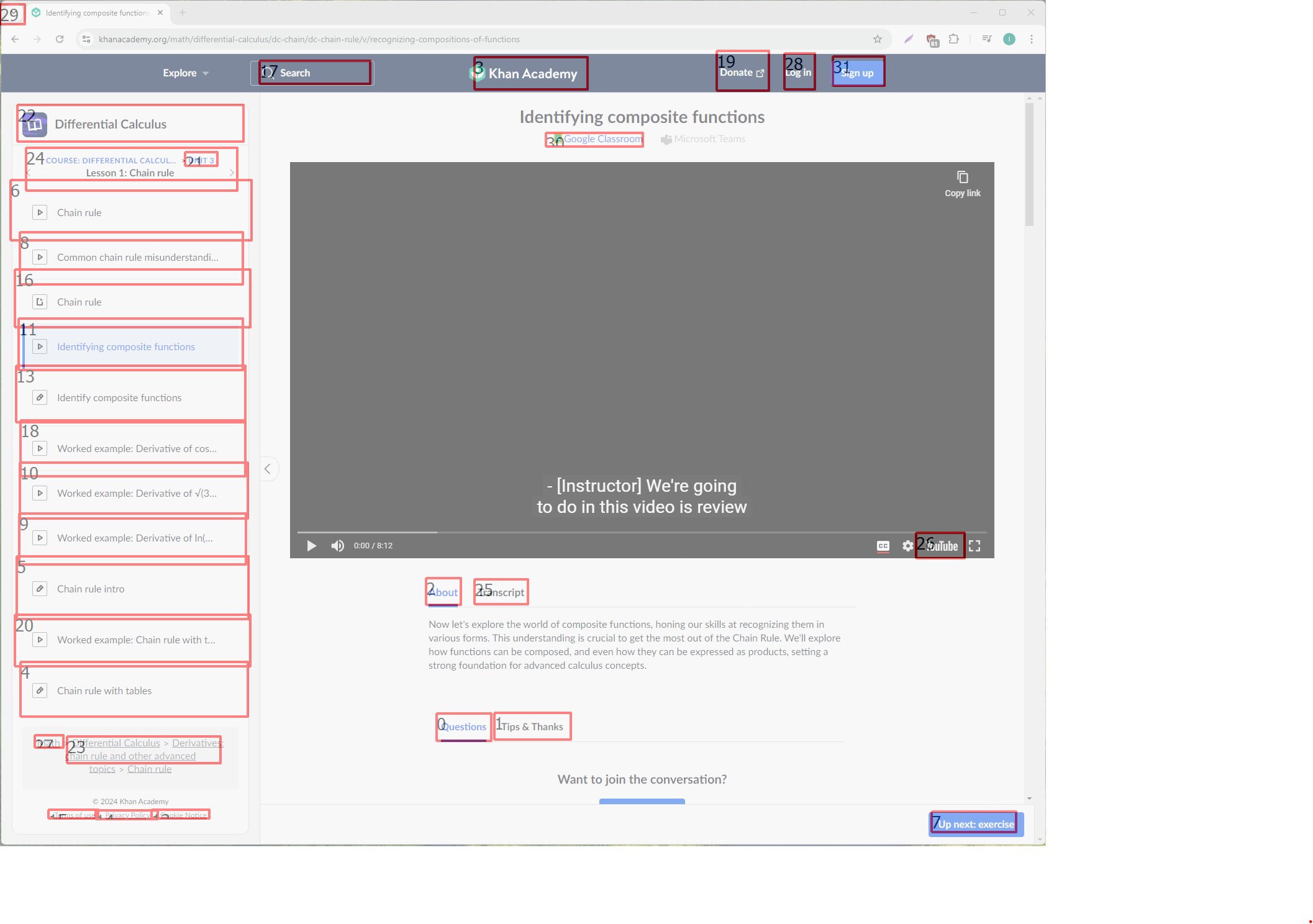}
    \caption{Example GUI navigation on \textit{KhanAcademy}, with detected interactables highlighted in red bounding boxes and numbered for hands-free control.}
    \label{fig:voice-example}
\end{figure}

We generated synthetic navigation workflows mimicking a real user experience by prompting ChatGPT to produce popular pages on \textit{KhanAcademy} and test actions that could be performed in those pages. From the answers, we formed a test set containing 15 screen-action pairs within 7 test cases. Given a trace of $N$ states, model success is quantified by the number of state steps the model can complete automatically without user interventions. We traversed the 7 test cases using speech-controlled hands-free navigation with the following 5 interactable detectors: (1) \textbf{Interactable Detector} (ours, $\S$\ref{section:interactabledetector}); (2) \textbf{web7k} or (3) \textbf{web350k} model trained on 7k noise-cleaned and 350k computer websites~\cite{webui}, respectively; (4) \textbf{VINS} model trained on 350k websites~\cite{webui} and fine-tuned on 4.8k Android screens~\cite{vins}; and the (5) \textbf{EasyOCR} OCR model~\cite{easyocr}. Notably, in this experiment we are down-sampling images to all detectors to a small resolution with the smaller screenshot dimension being $320$ pixels; this is to match the input configuration of the \textit{web7k} and \textit{web350k} models by Wu~\etal~\cite{webui}, for a fair comparison. Table \ref{tab:RQ1-results} presents the results of our experiment for speech-controlled GUI navigation of the 7 ChatGPT-generated test cases. An example of the user experience in the voice-controlled GUI navigation is shown in Figure \ref{fig:voice-example}; detected interactables are shown and numbered, so the user can select and click an interactable hands-free with the \texttt{click <number>} command.

\begin{figure}[h!]
    \centering
    \begin{subfigure}
        \centering
        \includegraphics[width=0.485\textwidth]{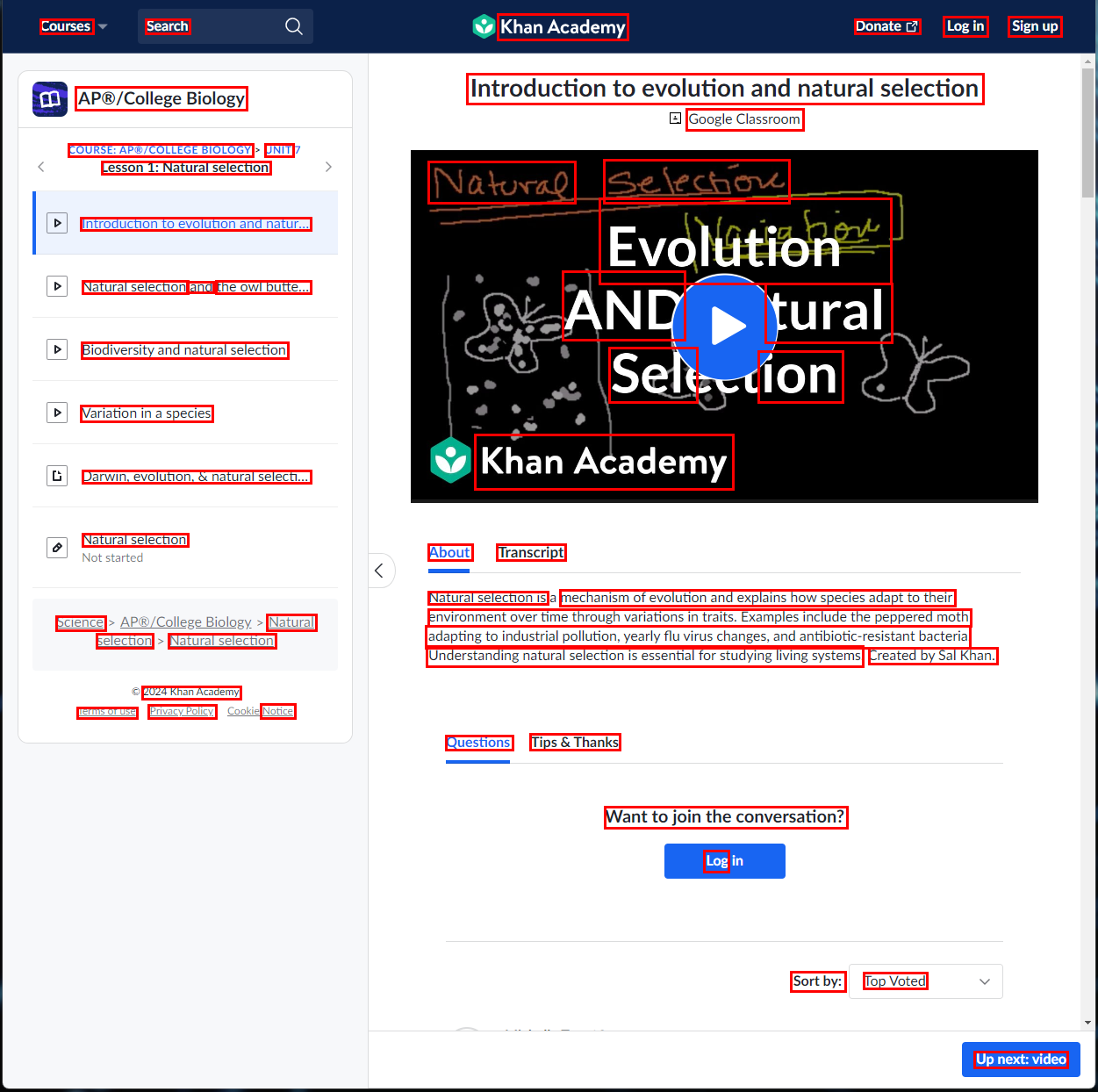}
    \end{subfigure}%
    \begin{subfigure}
        \centering
        \includegraphics[width=0.485\textwidth]{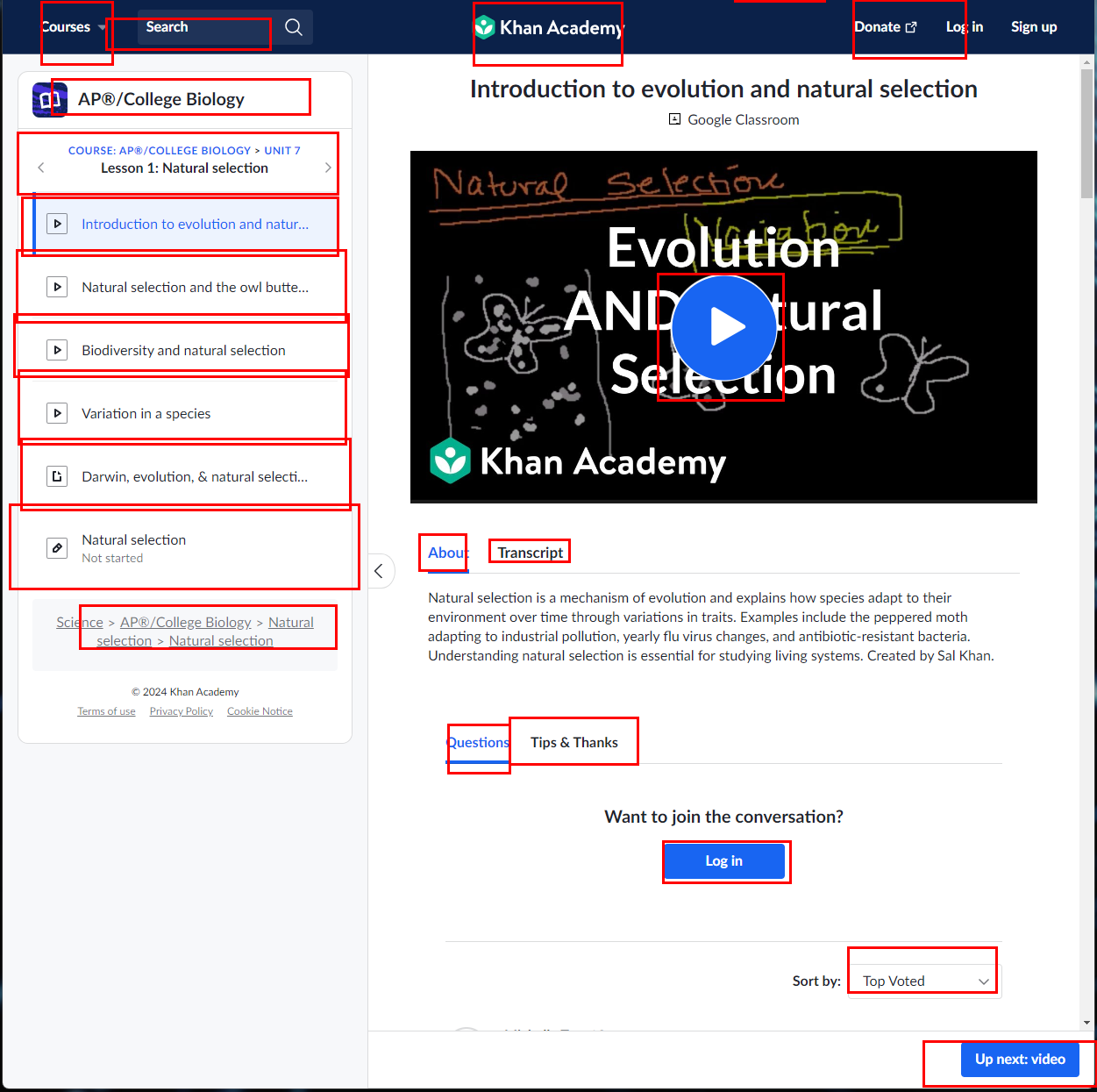}
    \end{subfigure}%
    \caption{Predicted Interactables in red, given by \textit{OCR} (left) and our \textit{Interactable Detector} at $320$-pixel resolution (right) for Test G.3. \textit{OCR} mislabels non-interactable text and misses the blue video play-button, which was the objective of Test G.3.}
    \label{fig:ocrvsinteract}
\end{figure}

Navigation using our \textit{Interactable Detector} outperforms the FCOS model in~\cite{webui} trained on \textit{web7k}, \textit{web350k} and \textit{VINS} by a large margin. Moreover, pure \textit{OCR} \cite{easyocr} outperforms our method in this experiment because all clicked interactables in the tests included text, except in Test G.3 (video play-button in Figure \ref{fig:ocrvsinteract}) -- this is the only interactable missed by the \textit{OCR} method. Compared to OCR, our detector is robust to text-free interactables. Further, \textit{OCR} incorrectly labels non-interactable text as valid interactables. Figure \ref{fig:ocrvsinteract} shows the non-cluttered and usable \textit{Interactable Detector} against \textit{OCR} in Test G.3.

    \begin{table*}[t]
        \centering
        \caption{Trace conversion rates for various models. * indicates a test case with the same end state as another test case, due to valid alternative navigation paths.}
        \begin{tabular}{r|c|ccc|c|c|ccc|ccc|ccc|c|}
        \multicolumn{1}{c|}{} &
          Test A &
          \multicolumn{3}{c|}{Test B} &
          Test C &
          Test D &
          \multicolumn{3}{c|}{Test E} &
          \multicolumn{3}{c|}{Test F} &
          \multicolumn{3}{c|}{Test G} &
          \multirow{2}{*}{Accuracy} \\
        \multicolumn{1}{c|}{} & 1 & 1 & 2 & 3 & 1 & 1 & 1 & 2 & 3 & 1 & 2 & 3 & 1 & 2 & 3 &  \\ 
        \textbf{Ours} ($320$-pixel)          & \textcolor{green}{\cmark} & \textcolor{red}{\xmark} & \textcolor{green}{\cmark}& \textcolor{red}{\xmark} & \textcolor{green}{\cmark}& \textcolor{red}{\xmark}  & \textcolor{green}{\cmark}& \textcolor{green}{\cmark}& \textcolor{green}{\cmark}& \textcolor{green}{\cmark} & \textcolor{red}{\xmark}  & \textcolor{red}{\xmark}  & \textcolor{green}{\cmark}& \textcolor{green}{\cmark} & \textcolor{green}{\cmark} & 67\% \\
        \textit{OCR}          & \textcolor{green}{\cmark}& \textcolor{green}{\cmark}& \textcolor{green}{\cmark}& \textcolor{green}{\cmark} & \textcolor{green}{\cmark} & \textcolor{green}{\cmark}  & \textcolor{green}{\cmark} & \textcolor{green}{\cmark} & \textcolor{green}{\cmark} & \textcolor{green}{\cmark}  & \textcolor{green}{\cmark}  & \textcolor{green}{\cmark}  & \textcolor{green}{\cmark} & \textcolor{green}{\cmark}  & \textcolor{red}{\xmark}  & 93\% \\
        \textit{web7k}        & \textcolor{red}{\xmark} & \textcolor{red}{\xmark} & \textcolor{red}{\xmark} & \textcolor{red}{\xmark} & \textcolor{red}{\xmark} & \textcolor{red}{\xmark}  & \textcolor{red}{\xmark} & \textcolor{red}{\xmark} & \textcolor{red}{\xmark} & \textcolor{red}{\xmark}  & \textcolor{red}{\xmark}  & \textcolor{red}{\xmark}  & \textcolor{red}{\xmark} & \textcolor{red}{\xmark}  & \textcolor{red}{\xmark}  & 0\%  \\
        \textit{web350k}      & \textcolor{red}{\xmark} & \textcolor{green}{\cmark} & \textcolor{red}{\xmark} & \textcolor{red}{\xmark} & \textcolor{green}{\cmark} & \textcolor{red}{\xmark}  & \textcolor{green}{\cmark} & \textcolor{red}{\xmark} & \textcolor{green}{\cmark} & \textcolor{red}{\xmark}  & \textcolor{red}{\xmark}  & \textcolor{red}{\xmark}  & \textcolor{red}{\xmark} & \textcolor{red}{\xmark}  & \textcolor{red}{\xmark}  & 27\% \\
        \textit{VINS}         & \textcolor{red}{\xmark} & \textcolor{green}{\cmark} & \textcolor{red}{\xmark} & \textcolor{red}{\xmark} & \textcolor{red}{\xmark} & \textcolor{red}{\xmark}  & \textcolor{red}{\xmark} & \textcolor{red}{\xmark} & \textcolor{red}{\xmark} & \textcolor{red}{\xmark}  & \textcolor{red}{\xmark}  & \textcolor{red}{\xmark}  & \textcolor{red}{\xmark} & \textcolor{red}{\xmark}  & \textcolor{green}{\cmark}  & 13\%
        \end{tabular}%
        \label{tab:RQ1-results}
    \end{table*}

\subsection{Trace Replication Across Platforms}
\label{section:gui-results-trace}

We demonstrate trace replication across different recording and replication Operating Systems. Specifically, we examine the test-cases E and G from $\S$\ref{section:gui-results-nav}, which our \textit{Interactable Detector} executed correctly. We record a trace by manually demonstrating tasks E (E.1-E.3) and G (G.1-G.3) on MacOS ($3072\times 1920$ resolution). For a larger test size, we also consider separate Traces X, Y, Z recorded on MacOS, which are very common single-action tasks to log in to an account (X), click on a learning item (Y), click on a topic unit (Z). We then conduct Trace Replication of the MacOS trace on a Windows 11 laptop ($1080\times 1920$ resolution); this demonstrates successful platform generalizability and independence despite GUI changes like OS and resolution. We assess performance using the proportion of states where the \SystemName automatically clicked on the correct target interactable.

Here, we again utilize the low-resolution $320$-pixel Interactable Detector. Also, because this input down-sampling leads to low-resolution FPN and Centerness feature maps, our Screen Similarity model is also operating at low-resolution. This enables us to record and replicate traces with real-time FPS, particularly on our MacOS CPU hardware; however, we trade off replication performance. Table \ref{tab:trace-replication} shows the number and proportion of states correctly \textit{replicated} on Windows 11 based on the MacOS-recorded traces.
        
\begin{table*}[h!]
    \centering
    \caption{Trace replication process with GUI changes (OS, pixel resolution, pop-up windows).}
    \begin{tabular}{r|ccc|ccc|c|c|c|c|}
      \multicolumn{1}{c|}{}        &
      \multicolumn{3}{c|}{Trace E} &
      \multicolumn{3}{c|}{Trace G} &
      \multicolumn{1}{c|}{Trace X} &
      \multicolumn{1}{c|}{Trace Y} &
      \multicolumn{1}{c|}{Trace Z} &
      \multirow{2}{*}{Total} \\
    \multicolumn{1}{c|}{} & 1 & 2 & 3 & 1 & 2 & 3 & 1 & 1 & 1 &  \\ 
    \cline{2-11}
    State Replicated & \textcolor{green}{\cmark} & \textcolor{red}{\xmark} & \textcolor{green}{\cmark} & \textcolor{red}{\xmark} & \textcolor{red}{\xmark} & \textcolor{red}{\xmark} & \textcolor{green}{\cmark} & \textcolor{green}{\cmark} & \textcolor{green}{\cmark} & $5/9$ \\
    Replication Accuracy & \multicolumn{3}{c|}{$66.6\%$} & \multicolumn{3}{c|}{$0\%$} & $100\%$ & $100\%$ & $100\%$ & $\mathbf{55.6\%}$ \\
    \end{tabular}%
    \label{tab:trace-replication}
\end{table*}

We show that our novel Action Matching mechanism for Trace Replication is deployable in a practical scenario, and robust to noise. The common mistake in replication is that the click position may be valid but not contained in noisy bounding-box detection of small clicked interactables (E.2). This is also the case in Trace G, which involved small target interactables with noisy predicted bounding boxes. This is because we are using our low-resolution ($320$-pixel) version of our Interactable Detector and Screen Similarity models. Further work is necessary to improve practicality by using our full-resolution models to better-detect small interactables.

\section{Conclusion}

We propose the novel end-to-end \SystemName system which has a) a suite of custom data-collection and auto-labeling algorithms, and b) an ML-based model for understanding and auto-navigating applications running on desktop computers (especially websites) and on Android phones. A test-time, \SystemName is independent of platform-specific APIs, unlike a significant number of eminent GUI Automators \cite{prefab,sikuli,sikuli-slides,sugilite,pumice,vasta,etna,actionshot}. \SystemName has excellent detection of interactables and state changes, relying only on GUI screenshots. A proof of concept system was shown to work on \textit{KhanAcademy}, to show how \SystemName can enable hands-free automated GUI traversal and Trace Replication, while being robust to GUI changes.

While OCR-based GUI understanding was shown to be supremely effective in our small trial, our expectation is that users will ultimately benefit from data and low-shot models that train on target-applications. Naturally, these too benefit from large-scale and general-purpose data collection, which we hope will be easier with our release of quite general-purpose code.

\begin{acks}
We acknowledge the \textit{JADE} High-Performance Computing facility owned by Oxford University and managed by the UK Hartree Institute.
\end{acks}

\bibliographystyle{ACM-Reference-Format}
\bibliography{references}

\onecolumn 

\appendix
\newpage
\section{FPN-Centerness Augmentation for Data Separability}
\label{appendix:centernessablation}

\begin{figure}[h!]
    \centering
    \includegraphics[width=0.8\linewidth]{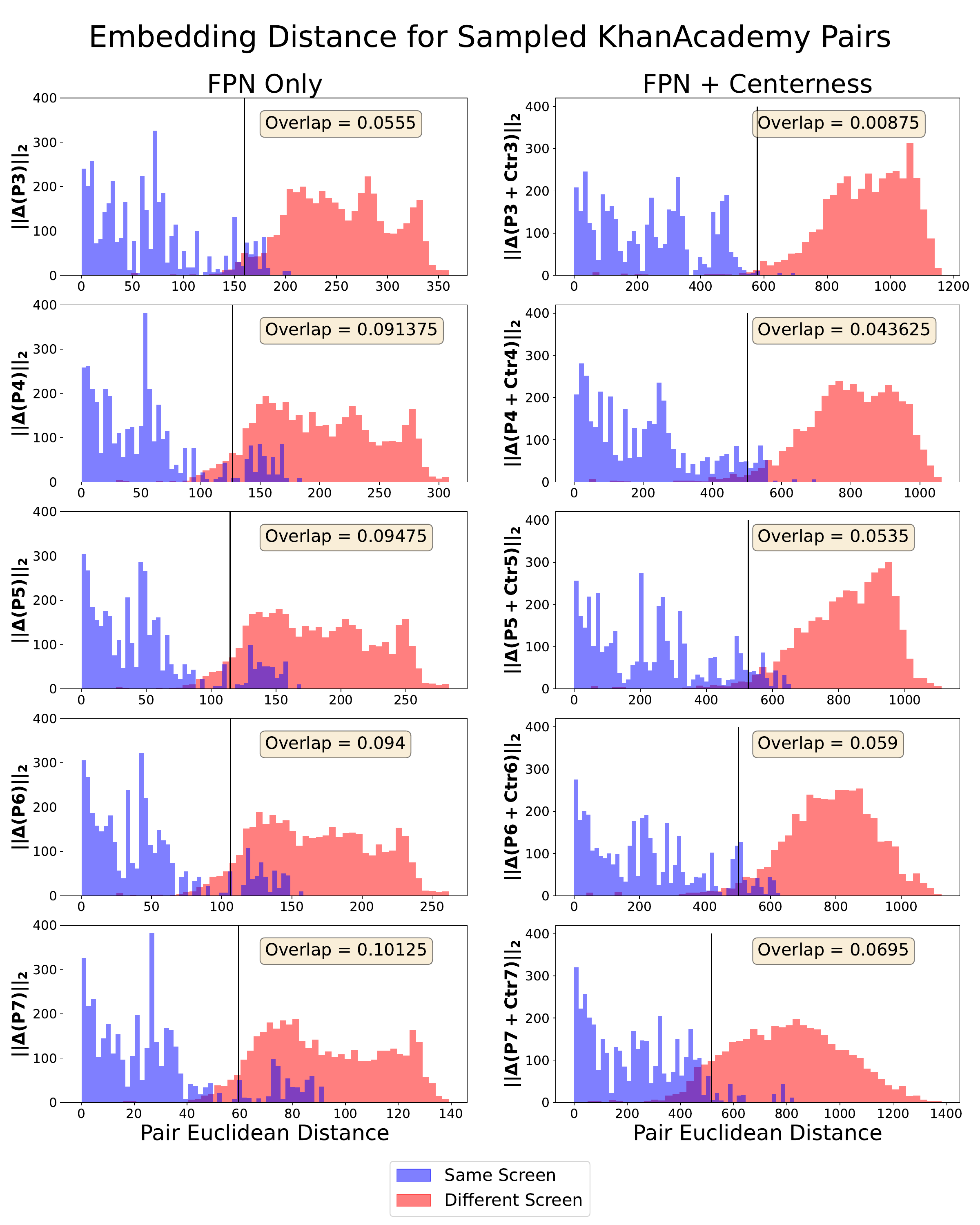}
    \caption{Distribution of the Euclidean Distance for 8000 Training-set pairs of \textit{KhanAcademy} GUI screenshots, in two cases: \textit{left} column is P3-P7 only; \textit{right} column is addition of normalized P3-P7 and corresponding Centerness 3 to Centerness 7 (short-hand \textit{Ctr3}-\textit{Ctr7}). The straight vertical line denotes the decision threshold.}
    \label{fig:screensim-pairs-all}
\end{figure}

Figure \ref{fig:screensim-pairs-all} displays the increased data separability when augmenting the FPN feature maps P3-P7 of the FCOS with the corresponding Centerness feature maps. This is valuable for the Screen Similarity model, as it enables a more distinct learnt latent transformation among same-state and different-state input pairs.

\section{Hyperparameter Configurations for GUI Automation ML Models}
\label{appendix:hparams}
The hyperparameter configuration for the \textit{Interactable Detector} FCOS model, trained with the SGD optimizer, is given in Table \ref{tab:interactable_hparam}.

\begin{table}[h]
    \begin{minipage}{\textwidth}
	\centering
	\caption{Hyperparameters of the \textit{Interactable Detector} model.}
	\begin{tabular}{l c}
        \toprule
        Hyperparameter & Configuration \\
        \midrule
        Learning Rate $\eta$ & $0.64$ \\
        Batch Size $B$ & $256$ \\
        Training Bit Precision & $32$-bit \\
        SGD Momentum & $0$ \\
        SGD Weight Decay & $0$ \\
	
	\bottomrule
	\label{tab:interactable_hparam}
	\end{tabular}
    \end{minipage}
\end{table}

The hyperparameter configuration for the \textit{Screen Similarity Model}, trained with the AdamW optimizer, is given in Table \ref{tab:screensim_hparam}. This configuration applies to all (FPN and FPN-Centerness) Siamese architectures presented in $\S$\ref{section:screen-sim}.

\begin{table}[h]
    \begin{minipage}{\columnwidth}
	\centering
	\caption{Hyperparameters of the \textit{Screen Similarity} model.}
	\begin{tabular}{l c}
        \toprule
        Hyperparameter & Configuration \\
        \midrule
        Learning Rate $\eta$ & $0.0128$ \\
        Batch Size $B$ & $64$ \\
        Negative Loss Margin $m_n$ & $0.5$ \\
        Positive Loss Margin $m_p$ & $0.2$ \\
        Training Bit Precision & $16$-bit \\	
	
    \bottomrule
	\label{tab:screensim_hparam}
	\end{tabular}
    \end{minipage}
\end{table}

\end{document}